\RequirePackage{ifpdf}
\documentclass[preprint, preprintnumbrs]{article}
\usepackage{jheppub}
\usepackage{xcolor}
\usepackage{amsmath}
\usepackage{epsfig}
\usepackage{caption}
\usepackage{subcaption}
\usepackage{comment}
\usepackage{hyperref}
\usepackage{soul}
\pdfoutput=1

\newcommand{\changelocaltocdepth}[1]{%
  \addtocontents{toc}{\protect\setcounter{tocdepth}{#1}}%
  \setcounter{tocdepth}{#1}%
}

\newcommand{\roughly}[1]{\mathrel{\raise.3ex\hbox{$#1$\kern-0.85em
\lower1ex\hbox{$\sim$}}}}

\def\nn{\nonumber}

\newcommand{\ket}[1]{|#1\rangle}
\newcommand{\bra}[1]{\langle #1 |}
\newcommand{\be}{\begin{equation}}
\newcommand{\bee}{\begin{equation}}
\newcommand{\ee}{\end{equation}}
\newcommand{\beea}{\begin{eqnarray}}
\newcommand{\eea}{\end{eqnarray}}
\newcommand{\bea}{\begin{eqnarray}}

\def\nott#1{\setbox0=\hbox{$#1$}                
   \dimen0=\wd0                                 
   \setbox1=\hbox{/} \dimen1=\wd1               
   \ifdim\dimen0>\dimen1                        
      \rlap{\hbox to \dimen0{\hfil/\hfil}}      
      #1                                        
   \else                                        
      \rlap{\hbox to \dimen1{\hfil$#1$\hfil}}   
      /                                         
   \fi}                                         %

\def\uxsl{\hbox{/\kern-.4000em$u$}}
\def\uxslsm{\hbox{\smaller/\kern-.5600em$u$}}
\def\pxpsl{\hbox{/\kern-.5000em$p$}}
\def\epssl{\hbox{/\kern-.5600em$\epsilon$}}
\def\delsl{\hbox{/\kern-.7000em$\nabla$}}
\def\lxpsl{\hbox{/\kern-.5600em$l$}}
\def\kxpsl{\hbox{/\kern-.5600em$k$}}
\def\qxpsl{\hbox{/\kern-.3900em$q$}}
\def\rint{{\rm int}}

\def\smath#1{\text{\scalebox{.85}{$#1$}}}
\def\sfrac#1#2{\smath{\frac{#1}{#2}}}

\def\llangle{\langle\hspace{-2mm}\langle}
\def\rrangle{\rangle\hspace{-2mm}\rangle}
\def\trA{{\hbox{Tr}_{\rm sys}}}
\def\trB{{\hbox{Tr}_{\rm env}}}

\def\pref#1{(\ref{#1})}
\def\exd{{\rm d}}
\def\ol#1{{\overline{#1}}}

\def\cH{{\cal H}}

\def\cL{{\cal L}}

\def\cO{{\cal O}}

\def\cV{{\cal V}}
\def\cW{{\cal W}}

\def\cZ{{\cal Z}}

\def\bfp{{\bf p}}

\def\bfx{{\bf x}}
\def\bfy{{\bf y}}

\def\bmk{{\boldsymbol{k}}}
\def\bmp{{\boldsymbol{p}}}
\def\bmq{{\boldsymbol{q}}}
\def\bmx{{\boldsymbol{x}}}

\def\mfg{{\mathfrak g}}

\def\UV{{\scriptscriptstyle U\hbox{\kern-0.1em}V}}
\def\PPN{{\scriptscriptstyle P\hbox{\kern-0.1em}P\hbox{\kern-0.1em}N}}
\def\MN{{\scriptscriptstyle M\hbox{\kern-0.1em}N}}
\def\MNP{{\scriptscriptstyle M\hbox{\kern-0.1em}N\hbox{\kern-0.1em}P}}
\def\KK{{\scriptscriptstyle K\hbox{\kern-0.1em}K}}
\def\SM{{\scriptscriptstyle S\hbox{\kern-0.1em}M}}
\def\EH{{\scriptscriptstyle E\hbox{\kern-0.1em}H}}

\def\QCD{{\scriptscriptstyle Q\hbox{\kern-0.1em}C\hbox{\kern-0.1em}D}}
\def\IR{{\scriptscriptstyle I\hbox{\kern-0.1em}R}}
\def\TEV{{\scriptscriptstyle T\hbox{\kern-0.1em}E\hbox{\kern-0.1em}V}}

\def\llangle{\langle\hspace{-2mm}\langle}
\def\rrangle{\rangle\hspace{-2mm}\rangle}
\def\trB{{\hbox{Tr}_{\rm env}}}

\def\Tr{{\rm Tr}}
\def\LE{{\scriptscriptstyle LE}}

\def\UV{{\scriptscriptstyle U\hbox{\kern-0.1em}V}}
\def\PPN{{\scriptscriptstyle P\hbox{\kern-0.1em}P\hbox{\kern-0.1em}N}}
\def\MN{{\scriptscriptstyle M\hbox{\kern-0.1em}N}}
\def\MNP{{\scriptscriptstyle M\hbox{\kern-0.1em}N\hbox{\kern-0.1em}P}}
\def\KK{{\scriptscriptstyle K\hbox{\kern-0.1em}K}}
\def\SM{{\scriptscriptstyle S\hbox{\kern-0.1em}M}}
\def\EH{{\scriptscriptstyle E\hbox{\kern-0.1em}H}}

\def\QCD{{\scriptscriptstyle Q\hbox{\kern-0.1em}C\hbox{\kern-0.1em}D}}
\def\IR{{\scriptscriptstyle I\hbox{\kern-0.1em}R}}
\def\TEV{{\scriptscriptstyle T\hbox{\kern-0.1em}E\hbox{\kern-0.1em}V}}
\def\aff{{a\hbox{\kern-0.1em}f\hbox{\kern-0.1em}f}}

\newcommand{\dd}{\mathrm{d}}
\newcommand{\Eq}[1]{Eq.~(\ref{#1})}
\newcommand{\Eqs}[1]{Eqs.~(\ref{#1})}
\newcommand{\Rea}{\Re \mathrm{e}\,}
\newcommand{\Ima}{\Im \mathrm{m}\,}
\newcommand{\Real}[1]{\Rea \left[ #1 \right] }
\newcommand{\Imag}[1]{\Ima \left[ #1 \right] }
\newcommand{\bs}[1]{\boldsymbol{#1}}
\newcommand{\Fig}[1]{Fig.~{\ref{#1}}}

\usepackage{wrapfig}

\title{   Does decoherence violate decoupling?   }

\author[a,b,c]{C.P.~Burgess,}
\author[d]{Thomas Colas,}
\author[e]{R.~Holman,}
\author[b,f,g]{Greg Kaplanek}

\affiliation[a]{Department of Physics \& Astronomy, McMaster University, Hamilton, ON, Canada, L8S 4M1}

\affiliation[b]{Perimeter Institute for Theoretical Physics, Waterloo, ON, Canada, N2L 2Y5}

\affiliation[c]{School of Theoretical Physics, Dublin Institute for Advanced Studies, 10 Burlington Rd., Dublin, Ireland}

\affiliation[d]{Department of Applied Mathematics and Theoretical Physics, University of Cambridge, Wilberforce Road, Cambridge, CB3 0WA, UK}

\affiliation[e]{Minerva University, 14 Mint Plaza, San Francisco, CA 94103, USA}

\affiliation[f]{Department of Physics, Imperial College, London, SW7 2AZ, UK}

\affiliation[g]{Department of Electrical Engineering and Computer Science, Syracuse University, NY 13210, USA}

\emailAdd{cburgess@perimeterinstitute.ca}
\emailAdd{tc683@cam.ac.uk}
\emailAdd{rh4a@andrew.cmu.edu}
\emailAdd{gkaplane@syr.edu}

\date{today}

\abstract{Recent calculations in both flat and de Sitter spacetimes have highlighted a tension between the decoupling of high-energy physics from low-energy degrees of freedom and the expectation that quantum systems decohere due to interactions with unknown environments. In effective field theory (EFT), integrating out heavy fields should lead to Hamiltonian time evolution, which preserves the purity of low-energy states.  This is consistent with the fact that we never observe isolated quantum states spontaneously decohering in the vacuum due to unknown high-energy physics.  However, when a heavy scalar of mass $M$ is traced out, the resulting purity of a light scalar with mass $m$ typically appears to scale as a power of $1/M$ (when $m\ll M$), an effect that \emph{cannot} be captured by a local effective Hamiltonian. We resolve this apparent paradox by showing that the purity depends on the resolution scale of the EFT and how the environment is traced out. We provide a practical method for diagnosing the purity of low-energy states consistent with EFT expectations, and briefly discuss some of the implications these observations have for how ultraviolet divergences can appear in decoherence calculations.}

\begin{document}
\maketitle

\section{Introduction}

Progress in science is possible because one is never required to study all degrees of freedom in the universe all at once. Physics in particular excels at focussing on a subset of variables in a simple limit to infer useful information about the behaviour of more complicated systems. Systems for which only a subset of variables are explicitly tracked are usually called {\it open systems}. Since essentially all scientific studies involve open systems it is surprising that open-system tools are not more widely used. 
 
In quantum mechanics not measuring a degree of freedom usually means marginalizing over its value (and so integrating it out). The emergence of novel phenomena like decoherence and thermalization show that the result after this has been done can behave dramatically differently than the initial system. The study of these effects falls into the category of open quantum systems. We follow some of the nomenclature of this field here, with the observed degrees of freedom generically called the `system' and the unobserved marginalized ones called the `environment'.

Effective field theories (EFTs) \cite{Weinberg:1978kz} are a particularly important framework for integrating out degrees of freedom in quantum systems in the special case that the ignored states are at high energy relative to the energies of interest in a physical process. In the modern understanding pretty much all of our successful theories are EFTs of this type. So one might expect that phenomena like decoherence and thermalization must be very common within the low-energy regime, since the EFT framework is at its heart an open system analysis. 

This expectation turns out to be wrong: EFTs lead to standard quantum behaviour for some choice of effective Hamiltonian with no exotic open effects like decoherence at all. That is just as well because we do not see isolated quantum states decohering around us all the time despite the fact that there are always very high energy degrees of freedom beyond reach of our measurements. Ultraviolet (UV) physics decouples from low-energy decoherence calculations.

There is a good reason for this that at its core relies on the fact that for EFTs the criterion separating the observed low-energy system from its unobserved high-energy environment involves the eigenvalues of a conserved quantity (energy). For Wilsonian EFTs defined by having low energy it is conservation of energy\footnote{Low energies are sometimes supplemented with other conserved things when defining a Wilsonian EFT -- such as a specific value of electric charge or baryon number -- but if so it is again important that these charges be conserved for avoiding things like decoherence. The same conclusions can also hold in time-dependent situations -- such as for cosmology -- provided the time-dependence is adiabatic \cite{Burgess:2003jk}.} that ultimately precludes EFTs from displaying the kinds of open-system non-Hamiltonian evolution that can cause things like decoherence \cite{EFTBook}. 

In this paper we test this general expectation by comparing it with the results of recent explicit decoherence calculations in which the time-evolution of the purity of a `system' field, $\sigma$, is computed when a second `environmental' field, $\phi$, is integrated out \cite{Burgess:2024eng, Colas:2024ysu}. We start by first spelling out the decoupling argument more explicitly, with \S\ref{sec:PurityTheorem} giving a more detailed argument why decoherence cannot appear when high-energy states are integrated out. 

A summary of the results of several explicit perturbative purity calculations is then given in \S\ref{sec:ConcreteExamples} (with some details provided in Appendix \ref{App:InteractionPicture}). For some of these the interactions are simple enough that an exact solution is possible. In the limit that the mass $M$ of the environmental field becomes very large these calculations should become well-described by an EFT framework and so their contribution to decoherence should be suppressed. 

In particular, general EFT arguments tell us that any effect that occurs at a fixed finite order in $1/M$ can be captured by an effective interaction within an appropriate effective Hamiltonian, $H_{\rm eff}$. But Hamiltonian evolution always takes pure states to pure states and so one expects that decoherence effects in such theories should be smaller than any finite power of $1/M$. This is {\it not} borne out by explicit calculations, which therefore contain a puzzle: how can having decoherence be proportional to any power of $1/M$ be consistent with the existence of a local Wilsonian Hamiltonian description? 

This paper aims to explain how these calculations can be consistent with the general arguments against decoherence in EFTs, without undermining Wilsonian methods, with the main argument given in \S\ref{sec:Decoupling}. Both the calculations and the general arguments are correct and this reveals a subtlety in interpreting purity calculations, particularly so far as the contribution of UV degrees of freedom are important. 

There are many ways to trace out UV degrees of freedom, particularly if the full system energy eigenstates are only approximately known. Although each correctly computes the purity evolution of low-energy states, they do so subject to different assumptions about precisely what the environment is that has been traced out. Only the specific calculation that accurately captures the effects of energy eigenstates provides a good diagnostic of the low-energy purity, and this agrees with the general expectation that pure states remain pure. \S\ref{sec:Decoupling} also describes a simple practical way to compute the purity evolution predicted by integrating out high-energy states, without having to know in advance precisely what the high-energy energy eigenstates are.

Because UV divergences are a specific type of UV behaviour our conclusions are also relevant to the discussion of whether divergences can appear in decoherence calculations (they can), and how they are renormalized if they do (we argue they can be renormalized in the well-known way appropriate for dealing with composite operators). A preliminary version of this discussion is given in \S\ref{sec:UVdivergences}. Finally \S\ref{sec:Conclusions} briefly summarizes our conclusions. Appendices \ref{App:SchrodingerPicture} and \ref{App:OpenEFT} provide equivalent and complementary discussions of our example calculations using the Schr\"odinger picture framework and using the Schwinger-Keldysh framework.

\subsection{Theorem: the purity of decoupling}
\label{sec:PurityTheorem}

In this section we argue that integrating out heavy physics always leads to Hamiltonian evolution (and so in particular cannot change a pure state into a mixed state). This is the benchmark against which we compare various explicit purity calculations in later sections.

In an operator language a low-energy theory is obtained by projecting onto a subspace of the Hilbert space defined by energy (or energy plus values of other conserved quantities). We assume that the only measurements of interest lie in the low-energy sector of states having energy $E < \Lambda$ where $\Lambda$ is an arbitrary UV cutoff for the theory. The projection operator onto the low-energy sector is $P_\Lambda$, defined for each energy eigenstate $H|E\rangle = E | E\rangle$ by:
\be
\begin{cases} \label{Eq:PLambda}
   \; P_\Lambda | E \rangle = 0 \quad & \hbox{if} \quad E > \Lambda \\
  \; P_\Lambda | E \rangle = | E \rangle \quad &\hbox{if} \quad E < \Lambda 
\end{cases} 
\ee
This satisfies $P^2_\Lambda = P_\Lambda$. In the low-energy theory the only observables considered have the form, 
\be
   \cO_\LE = P_\Lambda O \, P_\Lambda
\ee
for some hermitian $O$ and all allowed low-energy states $ | \Psi(t) \rangle_\LE := P_\Lambda | \Psi(t) \rangle$ satisfy 
\be
  P_\Lambda | \Psi(t) \rangle_\LE = | \Psi(t) \rangle_\LE \,.
\ee
In the Schr\"odinger picture the evolution of states is given in the full theory by
\be
   | \Psi(t) \rangle = U(t, t_0) | \Psi(t_0) \rangle = \exp\Bigl[ - i H(t-t_0) \Bigr] | \Psi(t_0) \rangle \,.
\ee
Evolution in the effective theory is then given by
\be
   | \Psi(t) \rangle_\LE =  P_\Lambda U(t,t_0) | \Psi(t_0) \rangle = U(t,t_0) P_\Lambda | \Psi(t_0) \rangle = P_\Lambda U(t,t_0) P_\Lambda | \Psi(t_0) \rangle_\LE \,,
\ee
which uses that $P_\Lambda$ commutes with $H$ and $P^2_\Lambda = P_\Lambda$. The evolution operator within the effective theory is therefore 
\be
   U_\LE(t,t_0) = P_\Lambda U(t,t_0) P_\Lambda = \exp\Bigl[ -i P_\Lambda H P_\Lambda (t-t_0) \Bigr] 
\ee
which is clearly Hamiltonian evolution, $U_\LE(t,t_0) = \exp[ -i H_{\rm eff} (t-t_0) ]$ with $H_{\rm eff} = P_\Lambda H P_\Lambda$. 

When using energy to integrate out states we are guaranteed to have Hamiltonian evolution. So it is an exact statement that pure states always remain pure within the effective theory as time goes by. The interesting question in this case is precisely what $H_{\rm eff}$ looks like when expressed in terms of low-energy fields $\phi_{\rm eff}(x) = P_\Lambda \phi(x) P_\Lambda$ (and the formalism of EFTs is aimed at identifying this as efficiently as possible -- see {\it e.g.}~\cite{EFTBook}). 

So far so good. But tracing out a sector is not simply a matter of projection. One must also incorporate the probability information that resides in the environmental part of the Hilbert space. To make the issues explicit, consider two uncoupled scalars,\footnote{Uncoupled scalars will be regarded as the leading approximation within a perturbative treatment of the more interesting case where the scalars are coupled.} a system field $\sigma$ with mass $m$ and an environmental field $\phi$ with mass $M \gg m$. The total Hilbert space is the product of the space for each field separately: $\mathcal{H} = \mathcal{H}_{\sigma} \otimes \mathcal{H}_{\phi}$. If $m \ll M$ projecting out energies $E < \Lambda$ with $m \ll \Lambda \ll M$ is related intuitively to tracing out the heavy field, but strictly speaking the image of the projector $P_\Lambda$ in this case includes states with an overlap with $\mathcal{H}_\phi$. An extra step is needed to completely reduce the theory to a single-field Hilbert space. 

\begin{figure}
\centering
\includegraphics[width=0.25\textwidth]{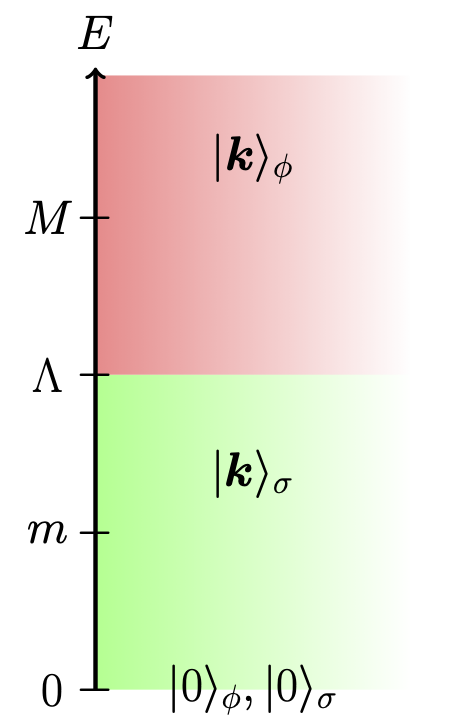}
\caption{Distribution of states in $\mathcal{H}:=\mathcal{H}_{\sigma} \otimes \mathcal{H}_{\phi}$ organized in terms of energy, with the red region corresponding to high energy states projected out ({\it i.e.} mapped to zero) via the low energy projection $P_{\Lambda}$ defined in Eq.~(\ref{Eq:PLambda}). The green region contains the low energy states preserved by the projection $P_{\Lambda}$. Note crucially that the $\phi$ vacuum $|0\rangle_{\phi}$ is in the low-energy sector.}
\label{fig:hierarchy}
\end{figure}

Let us assume that there are no initial correlations between $\sigma$ and $\phi$. Because of the mass hierarchy $m \ll \Lambda \ll M$ any state including at least one $\phi$ particlelike $\widehat{a}^{\dagger}_{\phi}(\bmk) \ket{0}_\phi  := |\bmk\rangle_\phi$  has energy $\omega_\phi \geq  M \gg \Lambda$, so the only low-energy $\phi$-sector state is the vacuum $\ket{0}_\phi$ defined by $\widehat{a}_{\phi}(\bmk) \ket{0}_\phi = 0$ for all $\bmk$ (see Figure \ref{fig:hierarchy}).
The only uncorrelated low-energy states therefore have the form
\be \label{LowEnergyStates}
   \rho = \varrho \otimes \Xi_\phi \quad \hbox{where} \quad
   \Xi_\phi = \ket{0}_\phi \, {}_\phi\bra{0} \,,
\ee
where $\varrho$ is a density matrix within the low-energy sector of $\cH_\sigma$. Tracing out high-energy states requires both a projection onto energies smaller than $\Lambda$ and the tracing out of the low-energy part of $\cH_\phi$ weighted by the vacuum state $\Xi_\phi$: $\llangle \,\cdots\, \rrangle = \trB[(\cdots) \,\Xi_\phi]$. Because ${}_\phi \langle 0 | 0 \rangle_\phi = 1$ we have $\rho^2 = \varrho^2 \otimes \Xi_\phi$ and so the purity of the full state $\rho$ is completely controlled by the purity of the state $\varrho$ within the $\sigma$ sector. The uniqueness of the low-energy state in the $\phi$ sector guarantees one can project onto the low-energy subspace defined by $\mathcal{H}_\sigma$ by tracing over $\mathcal{H}_\phi$ without loss of information.

The same reasoning also applies to interacting systems provided several key assumptions are valid. First we assume weak coupling so that the state of the system is only perturbatively different from the non-interacting factorized state \pref{LowEnergyStates}. Second, we assume the time evolution is adiabatic in the sense that it leaves the $\phi$ sector in its instantaneous vacuum state $\Xi_\phi(t)$.  Third, it is crucial that the projection $P_\Lambda$ is defined using only conserved quantities (low energy and perhaps the specification of the value of other conserved charges). 

These assumptions are required because interactions in general cause energy eigenstates to no longer factorize: $| \bmp, \bmq \rangle  \neq |\bmp \rangle_{\sigma} \otimes |\bmq \rangle_{\phi}$ for all times. The related requirements $\Lambda \ll M$ and adiabaticity help ensure that this factorization is maintained in practice for low-energy states, and so the process of tracing out the $\phi$ sector at time $t_0$ and evolving to time $t$ agrees with the process of evolving from $t_0$ to $t$ in the full theory and only then tracing out $\phi$. This distinction (in the presence of interactions) between $\phi$ and $\sigma$ sectors and energy eigenstates plays an important role in later sections, where use the name \textit{decoupled} basis for the energy eigenbasis and the name \textit{decohered} basis defined by tracing out the fields $\sigma$ and $\phi$. 

Notice that the adiabatic assumption also allows a deviation from the strict requirement that $P_\Lambda$ use only conserved quantities, in particular for situations with time evolving backgrounds for which fluctuation energy is not strictly conserved. In this case, it suffices for states to evolve adiabatically so that there are no quantum jumps to other states from an initial one. Note however if the interacting Hamiltonian is strongly time dependent, then a large number of $\sigma$ particles could be produced, taking states like those in eq.~(\ref{LowEnergyStates}) out of the low-energy space of states. This could potentially pose an issue in cosmological settings where the number of $\sigma$ particles created typically scales like $e^{aH/k}$ when $m \ll M$.

Summarizing the above requirements: we ask for a projection based on the eigenvalues of a conserved charge (notably energy) together with a scale hierarchy $M \gg m$ and adiabatic evolution in order to make it worthwhile to reduce the effective number of propagating degrees of freedom and to be able to restrict the heavy-field sector to a unique adiabatic vacuum. These are generic assumptions that are usually required for an effective field theory (EFT) treatment  -- see \cite{EFTBook} for details. We also assume a factorizable initial state with no cross-correlation between the two sectors, which is usually true of the perturbative ground state but is not in itself usually an EFT requirement.  Under these circumstances the purity of a low-energy state is not changed at all by the process of integrating out heavy physics, as is usually taken for granted when discussing low-energy effective Hamiltonians within an EFT framework. It is this general result that subsequent sections confront with explicit calculations for simple models.

\section{Concrete examples}
\label{sec:ConcreteExamples}

This section summarizes the results of several concrete purity calculations from the literature \cite{Burgess:2024eng, Colas:2024ysu} (and some new ones) that illustrate their dependence on particle masses. Thes examples are all variations on a basic theme, involving two scalar fields with masses $M > m$ (that can -- but need not -- be chosen to be hierarchically large) interacting in various ways. Their interactions are built using a Lagrangian of the form
\be
\cL = - \tfrac12 \Bigl[  (\partial \phi)^2 + (\partial \sigma)^2 +  M^2 \phi^2 +  m^2 \sigma^2   \Bigr] + \cL_\rint\,,
\ee
where the examples considered make different choices for the interaction term $\cL_\rint$. 

We examine cases that differ in whether $\cL_\rint$ is quadratic or cubic in the fields and in each case compare situations where $\cL_\rint$ contains zero or two derivatives (in order to contrast how the IR and UV properties of these two cases differ). The quadratic models are simplest because they are purely gaussian, involving two fields that mix by a small amount in the mass terms or kinetic terms, and so can be exactly solved. But gaussian models are also rarely representative of interacting theories so we also consider examples where this mixing arises through cubic interactions, for which we restrict for simplicity to interactions with the symmetry $\phi \to - \phi$. We summarize the main points these examples raise for decoupling in \S\ref{ssec:DecovsDeco}.

In all of our examples the interaction Hamiltonian obtained from $\cL_{\rm int}$ has the form
\be
   \cH_{\rm int}(x) = \sum_i \cO^i_{\rm sys}(x) \otimes \cO^i_{\rm env}(x) 
\ee
where operators $\cO_{\rm sys}^i$ act in the $\sigma$-field sector and $\cO^i_{\rm env}$ act in the $\phi$-field sector. Furthermore in our examples $\cO_{\rm sys}^i$ is linear in $\sigma$ and this simplifies the purity calculation because it can be computed at leading order separately for each momentum mode \cite{Burgess:2024eng, Burgess:2022nwu} provided the initial state is prepared with different momentum modes in uncorrelated states (such as in the vacuum). For the purposes of our leading-order discussion the reduced density matrix for the $\sigma$ field can be regarded as a product of density matrices for each mode, 
\be 
   \varrho := \hbox{Tr}_\phi \, \rho =  \bigotimes_{\bmk}  \varrho_{\bmk}  
\ee
where $\rho$ is the density matrix of both system and environment, and the purity is completely characterized by computing $\gamma_\bmk := \hbox{Tr}_\sigma \, \varrho^2_\bmk$ separately for each $\bmk$. 

For each choice for $\cL_{\rm int}$ we give the leading-order result for $\gamma_\bmk$ obtained when the field $\phi$ is traced out. At leading order in perturbation theory the prediction for $\gamma_\bmk$ is (see eq.~\pref{AppPurityMasterEqFNested})
\be \label{PurityMasterEqFNested}
    \gamma_\bmk    =  \gamma_\bmk(t_0) - 2 \sum_{ij} \int_{t_0}^t \exd s' \int_{t_0}^t \exd s  \; \hbox{Re}\Bigl[ W^{ij}_\bmk(s',s) \, \cW^{ji}_{-\bmk}(s',s) \Bigr] \,,
\ee
where 
\be
\label{eq:Wightmann:deft}  
   \int \frac{ \exd^3k}{(2\pi)^3} \; \cW^{ij}_{\bf{k}}(t,s) \, e^{ i \bf{k} \cdot (\bf{x} - \bf{y})}  = \llangle \, \cO^i_{\rm env}(t,\bfx) \, \cO^j_{\rm env}(s,\bfy) \,\rrangle - \llangle \, \cO^i_{\rm env}(t,\bfx) \, \rrangle \, \llangle \, \cO^j_{\rm env}(s,\bfy) \, \rrangle   \,. 
\ee
where, as above, $\llangle \, \cdots\,  \rrangle$ denotes tracing over the $\phi$ sector, and
\be \label{Wsysdeft}  
   \int \frac{ \exd^3k}{(2\pi)^3} \; W^{ij}_{\bf{k}}(t,s) \, e^{ i \bf{k} \cdot (\bf{x} - \bf{y})}
   =  \langle \cO^i_{\rm sys}(t,\bfx) \, \cO^j_{\rm sys}(s,\bfy) \rangle  -  \langle \cO^i_{\rm sys}(t,\bfx) \rangle \, \langle \cO^j_{\rm sys}(s,\bfy) \rangle  \,,
\ee
where $\langle \, \cdots \rangle$ denotes the expectation in the $\sigma$ sector.

This section simply quotes these results, with calculational details given in Appendix \ref{App:InteractionPicture}. We do, however, emphasize an important subtlety in these calculations - particularly when evaluating the real part, as instructed in \pref{PurityMasterEqFNested} -- that plays a role in later discussions. The subtlety arises because the time argument $t-s$ of the functions $W_\bmk^{ij}(t,s)$ and $\cW^{ij}_\bmk(t,s)$ is not real. It is instead evaluated with a small negative imaginary part that is taken to zero only at the end of the calculation. This imaginary part arises because the correlation functions are evaluated by summing over a complete resolution of unity between the two operators, and a negative imaginary part is required in order for this sum to be well-defined. 

\subsection{Mixing examples}\label{sec:Mixing}

We start with two examples where system and environment communicate only through a quadratic mixing interaction (so the model is at heart gaussian).

\subsubsection{Mass mixing} \label{ssec:MassMixing}

First consider mass mixing, for which
\be
   \cL_\rint = -  \mu^2 \, \phi \, \sigma \,.
\ee
This system can be solved exactly by diagonalizing the mass matrix. The two mass eigenstates in this case, $\phi_\pm$, have squared masses 
\be\label{masseigenvalues}
   M^2_\pm = \frac12 \Bigl[ (m ^2+M ^2) \pm \sqrt{(m ^2-M ^2)^2 + 4\mu^4} \Bigr]  \,,
\ee
which are both positive when $0 \leq \mu^2 \leq M  m $ (as we henceforth assume). The rotation that takes $(\sigma, \phi)$ to $(\phi_-, \phi_+)$ is given explicitly in eq.~\pref{Appmassmixingeigenbasis}. 

The perturbative limit in this case assumes $\mu^2 \ll M^2 - m^2$, in which case
\be\label{masseigenvalues+-}
   M^2_+  \simeq M^2 + \frac{ \mu^4}{M^2 - m^2}    \quad \hbox{and} \quad
   M^2_-  \simeq m^2 -  \frac{ \mu^4}{M^2 - m^2}   \,, 
\ee
while the redefinition \pref{Appmassmixingeigenbasis} reduces to 
\be\label{DiagonalizationRedef}
  \phi \simeq \phi_+ - \frac{\mu^2 \phi_- }{M^2-m^2}  
  \quad \hbox{and} \quad
  \sigma \simeq \phi_- + \frac{\mu^2 \phi_+}{M^2-m^2}      \,.
\ee

We next compute the purity for the field $\sigma$ once the field $\phi$ is traced out, doing so perturbatively in powers of $\mu^2$ following \cite{Burgess:2024eng}. In this case inspection of $\cH_{\rm int} = - \cL_{\rm int}$ shows that $\cO_{\rm sys} = \sigma$ and $\cO_{\rm env} = \mu^2 \phi$, so if the two fields are prepared in their vacuum state the Wightman functions evaluate to
\be \label{FlatWk}
 W_{\bmk}(t,s) =  \frac{1}{2 \omega_\sigma(k)} \, e^{-i\omega_\sigma(k)(t-s)} \quad \hbox{and} \quad
 \cW_{\bmk}(t,s) =  \frac{\mu^4}{2 \omega_\phi(k)} \, e^{-i\omega_\phi(k)(t-s)} \,,
\ee
where $\omega_\phi(k) = \sqrt{k^2 + M^2}$ and $\omega_\sigma(k) = \sqrt{k^2 + m^2}$ are the appropriate dispersion relation for each field.

If the interaction term is regarded as turning on instantaneously at $t=t_0$ (the `sudden' approximation) then the subsequent purity of the $\sigma$ state evaluates to  
\be  \label{PurityMasterEqFflat}
  \gamma_\bmk^{(s)}(t)   =  1  -  \frac{2\mu^4}{\omega_\sigma \omega_\phi(\omega_\sigma +\omega_\phi)^2} \, \sin^2 \Bigl[\tfrac12 (\omega_\sigma +\omega_\phi)(t-t_0) \Bigr]  
  \simeq  1  -  \frac{2\mu^4}{\omega_\sigma M^3} \, \sin^2 \Bigl[\tfrac12 M (t-t_0) \Bigr]  \,,
\ee
where the approximate equality applies if $k,m \ll M$. Alternatively, if the interaction is instead turned on only adiabatically in the remote past then one instead finds a time-independent result\footnote{See Appendix \ref{App:InteractionPicture} for a discussion of why this result for $\gamma_\bmk -1$ is half as large as the time-averaged version of \pref{PurityMasterEqFflat}.}
\be \label{PurityMasterEqFflatsecular}
  \bar \gamma_\bmk^{(a)}(t)  =  1  -  \frac{\mu^4}{2\omega_\sigma \omega_\phi(\omega_\sigma +\omega_\phi)^2}  
   \simeq  1  -  \frac{\mu^4}{2\omega_\sigma M^3}  \,.
\ee

\subsubsection{Kinetic mixing}
\label{ssec:KineticMixing}

We next consider a different kind of energy dependence for mixing, for which off-diagonal mixing between $\phi$ and $\sigma$ occurs only in terms of differentiated fields, 
\be
   \cL_\rint = -  \xi \, \partial_\mu \phi \, \partial^\mu \sigma \,,
\ee
with $|\xi| < 1$. 

This is again a gaussian system since the kinetic and mass terms can be simultaneously diagonalized using the transformation \pref{Appkinmixingeigenbasis} with mixing angle given in \pref{Appmassmixingeigenbasiskin}. The resulting mass eigenvalues are
\be\label{masseigenvalueskin}
   M^2_\pm = \frac{1}{2(1-\xi^2)} \Bigl[ (m ^2+M ^2) \pm \sqrt{ (M^2 - m^2)^2 + 4 \xi^2 M^2 m^2} \Bigr]  \,.
\ee
For small $\xi$ these reduce to
\be\label{kinmasseigenvalues+-}
   M^2_+  \simeq M^2 \left[ 1  + \frac{ \xi^2 ( M^2 + m^2)}{M^2 - m^2}  \right]  + \cO(\xi^4) \quad \hbox{and} \quad
   M^2_-  \simeq m^2 \left[ 1  - \frac{ \xi^2 ( M^2 + m^2)}{M^2 - m^2}  \right]   + \cO(\xi^4)   \,,
\ee
and the transformations that diagonalize the Lagrangian simplify to
\be\label{DiagonalizationRedefxi}
  \phi  \simeq  \phi_+ + \frac{\xi  m^2 \phi_-}{M^2 - m^2}   + \cO(\xi^2)   \quad  \hbox{and} \quad
  \sigma  \simeq  \phi_- - \frac{\xi  M^2 \phi_+}{M^2 - m^2}  + \cO(\xi^2)   \,. 
\ee

We wish to compute the purity for the field $\sigma$ once the field $\phi$ is traced out, with both fields prepared in their vacuum state. To this end Appendix \ref{App:InteractionPicture} computes the Hamiltonian and shows that $\cH_{\rm int}$ is built from the operators $\cO^\mu_{\rm sys} = \partial^\mu \sigma$ and $\cO^\nu_{\rm env} = \xi \, \partial^\mu \phi$, so the required Wightman functions are
\be \label{FlatWk2}
 W^{\mu\nu}_{\bmk}(t,s) =  \frac{k^\mu k^\nu}{2 \omega_\sigma(k)} \, e^{-i\omega_\sigma(k)(t-s)} \quad \hbox{and} \quad
 \cW^{\mu\nu}_{\bmk}(t,s) =  \frac{\xi^2 k^\mu k^\nu}{2 \omega_\phi(k)} \, e^{-i\omega_\phi(k)(t-s)} \,,
\ee
where $k^0 = \omega_\sigma(k)$ in $W^{\mu\nu}_\bmk$ and $k^0 = \omega_\phi(k)$ in $\cW^{\mu\nu}_\bmk$.  
 
Using this in eq.~\pref{PurityMasterEqFNested} then leads to the `sudden approximation' result
\be  \label{PurityMasterEqFflatkin}
  \gamma_\bmk^{(s)}(t)   =  1  -  \frac{2\xi^2 (\omega_\sigma \omega_\phi + k^2)^2}{\omega_\sigma \omega_\phi(\omega_\sigma +\omega_\phi)^2} \, \sin^2 \Bigl[\tfrac12 (\omega_\sigma +\omega_\phi)(t-t_0) \Bigr]  
  \simeq  1  -  \frac{2\xi^2\omega_\sigma}{ M} \, \sin^2 \Bigl[\tfrac12 M (t-t_0) \Bigr]  \,,
\ee
where the approximate equality takes $M$ much larger than $m$ and $k$. The corresponding answer in the adiabatic approximation is instead
\be \label{PurityMasterEqFflatsecularkin}
   \gamma_\bmk^{(a)}(t) = 1 - \frac{\xi^2 (\omega_\sigma \omega_\phi + k^2)^2}{2\omega_\sigma \omega_\phi(\omega_\sigma +\omega_\phi)^2} \simeq  1  -   \frac{\xi^2\omega_\sigma}{ 2M} \,.
\ee

\subsection{Cubic interactions}

We next describe two examples where $\phi$ and $\sigma$ couple through a {\it bona fide} cubic interaction, considering first a cubic coupling in the scalar potential and then where the cubic coupling involves derivative interactions. These examples provide evidence for how less trivial system/environment interactions can generate $\gamma \neq 1$ when treated perturbatively. In all cases we consider only interactions that preserve the symmetry $\phi \to - \phi$ (something that was not possible for the quadratic mixing examples discussed earlier).

\subsubsection{Nonderivative cubic couplings}

The minimal interaction coupling the two fields in the scalar potential that is invariant under $\phi \to - \phi$ is cubic, with
\be
\cL_{\rm int} = - \mfg \, \phi^2 \sigma 
\ee
for coupling constant $\mfg$ with dimensions of mass (in fundamental units). 

The operators appearing in the interaction Hamiltonian at lowest order now are $\cO_{\rm sys} = \sigma$ and $\cO_{\rm env} = \mfg \, \phi^2$. This ensures $W_\bmk$ is again given by \pref{FlatWk} but the relevant environmental Wightman function instead becomes \cite{Burgess:2024eng}
\be  \label{FlatMassivecWk}
\cW_{\bf{k}}(t,s)
 =   \frac{i\mfg^2}{2(2\pi)^2k(t-s)} \int_0^\infty \frac{p \, \exd p}{\omega _\phi}\, e^{-i \omega _\phi (t-s)} \left[ e^{-i \omega _+(t-s)} - e^{-i \omega _-(t-s)} \right] \,,
\ee
with $\omega_\pm := \sqrt{(p \pm k)^2 + M^2}$. Notice that the small negative imaginary part that is required of $t-s$ in the Wightman function ensures the convergence of the $p$ integration. 

In this case it is the differential version of \pref{PurityMasterEqFNested} that we evaluate. Tracing out $\phi$ with both fields prepared in the vacuum then gives the following rate of change for the purity
\be \label{PurityChangeMasterEqFSum2}
  \partial_t \gamma_\bmk = - \frac{2}{\omega_\sigma}  \int_{t_0}^t \exd s  \; \hbox{Re}\left[  e^{-i\omega_\phi (t-s)} \cW_{-\bmk}(t,s) \right] \,,
\ee
where $\cW_\bmk$ is given in \pref{FlatMassivecWk}. In the limit $k,m \ll M$ the purity evolution therefore becomes
\be \label{PurityChangeMasterEqFSum2x}
  \partial_t \gamma_\bmk \simeq  \frac{\mfg^2}{4\pi^2\omega_\sigma} \int_1^\infty \frac{\exd u}{u^2} \sqrt{u^2-1} \; \hbox{Im}\left[  e^{-2i u M (t-t_0)} \right] \,.
\ee

Integrating this in the `sudden' approximation then gives an expression for the oscillating purity in terms of Struve and Bessel functions \cite{Burgess:2024eng}. What is more useful here is the result when the coupling is adiabatically turned on in the remote past. This gives the time-independent result\footnote{This is again half as large as the time-averaged `sudden' result quoted in \cite{Burgess:2024eng}.}
\be \label{PurityMassiveEnvEqFSum2xsecular}
  \gamma_\bmk^{(a)} \simeq
  1  -   \frac{\mfg^2}{16\pi^2\omega_\sigma M} \int_1^\infty \frac{\exd u}{u^3} \sqrt{u^2-1}  = 1  -   \frac{\mfg^2}{64\pi\omega_\sigma M}  \,,
\ee

\subsubsection{Derivative cubic couplings}
\label{sec:DerivativeCubic}

Our final example assumes the interaction
\begin{eqnarray}\label{dim5Int}
 \cL_{\rm int} = - \tfrac12  \kappa \,  \partial_\mu \phi \,  \partial^\mu \phi \, \sigma \,,
\end{eqnarray}
where $\kappa$ is a coupling with dimension length (in fundamental units). To leading order in $\kappa$ this leads to an interaction Hamiltonian with $\cO_{\rm sys} = \sigma$ and $\cO_{\rm env} = \frac12 \, \kappa \, \partial_\mu \phi \, \partial^\mu \phi$. 

For a state initially prepared in the vacuum $W_\bmk$ is again given by \pref{FlatWk} but the environmental Wightman function in this case is
\bea \label{cWcubicderiv}
 \cW_{\bmk}(t_1,t_2) & = & \tfrac14\kappa^2 \int \exd^3 x \; \bigg[ \langle 0 | (\partial\phi)^2(t_1,\bmx) (\partial\phi)^2(t_2,\mathbf{0}) | 0 \rangle - \langle 0 | (\partial\phi)^2(t_1,\bmx)  | 0 \rangle \langle 0 | (\partial\phi)^2(t_2,\mathbf{0}) | 0 \rangle \bigg] \; e^{-i \bmk \cdot \bmx} \nn\\
& = &   \frac{\kappa^2}{ (8 \pi )^5 k} \int_{k}^\infty \exd P \int_0^k \exd Q\; \frac{ (P^2 - Q^2) \left( P^2 + Q^2 - 2k^2 - 4 \omega_+\omega_-\right)^2 }{ \omega_+ \omega_- } e^{ - i \left[ \omega_+ + \omega_- \right] (t_1 - t_2 )} \,,
\eea
where $\omega_\pm := \sqrt{\frac14(P\pm Q)^2 + M^2}$.

Inserting these expressions into the perturbative purity and assuming the interaction is turned on adiabatically in the remote past we get the time-independent result
\begin{equation} \label{PP3}
\gamma_{\bmk} = 1 - \frac{ \kappa^2}{  (8\pi)^5 k \omega_\sigma } \int_k^\infty \exd P \int_0^k \exd Q\; \frac{ (P^2 - Q^2) \left( P^2 + Q^2 - 2k^2 - 4 \omega_+ \omega_- \right)^2 }{ \omega_+ \omega_- \big( \omega_\sigma + \omega_+ + \omega_- \big)^2 } 
\end{equation}
where $\omega_\sigma = \omega_\sigma(k)$ and the $P$ integral turns out to converges in the UV. In the decoupling limit, for which $M \gg k,m$, this is well-approximated by
\begin{equation} \label{PPF}
\gamma_{\bmk} \simeq 1 - \frac{ \kappa^2 M^3 }{ 1024 \pi^4 \omega_{\sigma} }  \,.
\end{equation}
Although the appearance of $M$ in the numerator of this expression is at first sight unnerving, it must be kept in mind that use of a dimension-5 interaction like in \pref{dim5Int} only makes sense in the limit that $\kappa M \ll 1$, so if $\kappa^{-1} = \Lambda$ is a UV scale then consistency of the EFT interpretation requires $M \ll \Lambda$\footnote{To see this in an example, one can consider a simple partial UV completion in which the scalar $\sigma$ and $\phi$ couple to a heavier scalar $H$ with mass $\kappa^{-1} =\Lambda$, through an interaction of the form $(g_1 \sigma + g_2 \phi^2 )H$. Integrating out $H$ at tree level gives rise to many operators including $g_1 g_2 \sigma (\Box + \Lambda^2)^{-1} \phi^2$. Expanding out to dimension-5 and applying the equations of motion for $\phi$ one finds that this simplifies to a linear combination of $g_1 g_2 \kappa^2 \sigma (1 + \kappa^2 M^2) \phi^2$ and $g_1 g_2 \kappa^4 \sigma (\partial \phi)^2$. In order for the derivative expansion to be under control one clearly needs $\kappa M = M/\Lambda \ll 1$}. We return to the sensitivity to UV physics in \S\ref{sec:UVdivergences} where we discuss the related issue of UV divergences within purity calculations. 

\section{Decoupling and the $\epsilon$ prescription}
\label{sec:Decoupling}

The puzzle in expressions like \pref{PurityMasterEqFflatsecular}, \pref{PurityMasterEqFflatsecularkin}, \pref{PurityMassiveEnvEqFSum2xsecular} and \pref{PPF} is why $\gamma_\bmk$ differs from unity at all, since the arguments of \S\ref{sec:PurityTheorem} tell us that it is an exact result that the low-energy field prepared in a pure state does not decohere at all when a heavy field (prepared in its vacuum) is integrated out. On one hand, we see in \S\ref{sec:PurityTheorem} that \emph{integrating out} heavy degrees of freedom -- in the sense of projecting out degrees of freedom using conserved quantities (like energy or charge) -- always leads to Hamiltonian evolution and so any initially pure state always remains pure.  On the other hand, the explicit calculations in \S\ref{sec:ConcreteExamples} show that \emph{tracing out} massive fields causes the purity to evolve. 

It is true that expressions \pref{PurityMasterEqFflatsecular}, \pref{PurityMasterEqFflatsecularkin} and \pref{PurityMassiveEnvEqFSum2xsecular} imply $\gamma_\bmk - 1$ tends to zero as $M \to \infty$, however even this seems not quick enough. After all, decoupling implies that the virtual effects at low energies of the heavy field can be captured at low energies by a local Wilson action (or Hamiltonian) at any finite order in $1/M$. But {\it any} Hamiltonian evolution brings pure states to pure states and so ensures $\gamma_\bmk$ is unity for all time if $\gamma_\bmk(t_0) = 1$ for any particular time. So why isn't $\gamma_\bmk - 1$ smaller than any finite power of $1/M$?

This section describes how these two things can be reconciled, and how one identifies the correct decoherence calculation to perform -- {\it i.e.}~ the one that provides a good diagnostic of low-energy system purity -- when integrating out heavy states in practice, given only approximate information about the exact energy eigenstates and spectrum.

\subsection{Decoherence vs decoupling}
\label{ssec:DecovsDeco}

For the mixing examples of \S\ref{sec:Mixing} the absence of decoherence would have been explicit if the same calculation had been performed by integrating out the heavy mass eigenstate $\phi_+$ rather than $\phi$. Because the fields $\phi_\pm$ do not interact with one another in these examples it is obvious that integrating out $\phi_+$ cannot alter the purity of any state in the $\phi_-$ sector if these two sectors are initially uncorrelated (as is true in particular for their vacua). Initially pure $\phi_-$ states therefore always satisfy $\gamma_\bmk(t) = 1$, so why do the calculations of \S\ref{sec:Mixing} say otherwise?

For systems interacting only through mixing the answer is clear: the nonzero results \pref{PurityMasterEqFflatsecular} and \pref{PurityMasterEqFflatsecularkin} arise because of a misalignment between the projections in the two calculations -- the $\phi$ sector (or `decohered' sector) is related to the high-energy $\phi_+$ sector (or `decoupled' sector) by the rotation \pref{DiagonalizationRedef} or \pref{DiagonalizationRedefxi}, that becomes nontrivial at order $1/M^2$ and/or order $\xi$.  Expressions \pref{PurityMasterEqFflatsecular} and \pref{PurityMasterEqFflatsecularkin} correctly show how purity evolves when the field $\phi$ is traced out, but the existence of mixing implies the $\phi$ sector is not precisely the same as the high-energy sector. These nonzero results are the answer to a specific question -- what happens when the field $\phi$ is traced out -- and this is not a good diagnostic for what is seen at low energies when the heavy mass eigenstate $\phi_+$ is integrated out.

A similar alignment issue also arises in the case of a cubic interaction because in this case the interactions themselves imply that the $\phi$ operator does not precisely agree with the heavy mass-eigenstate field. But in this case there is no redefinition that completely removes the interactions and so likely also no redefinition that completely removes the decoherence to all orders in $1/M$.

The purity evolution depends on the details of the projection that removes the environment. When assessing the purity of a low-energy state after integrating out high-energy physics it doesn't matter that tracing out $\phi$ gives a nonzero result (though it can matter for other questions). The physical question of whether the system state remains pure after integrating out heavy physics instead depends on whether there {\it exists} a projection that gives Hamiltonian evolution (and so preserves the purity of low-energy states). 

When the environment is defined in terms of a conserved quantity the arguments of \S\ref{sec:PurityTheorem} tell us that such a projection does exist. It is less clear how to identify the projection that removes the illusion of purity change in the case where the full theory is not gaussian (or otherwise exactly solvable), short of performing an order-by-order diagonalization of the Hamiltonian to identify the precise energy eigenstates. How can one know in advance if a given purity calculation provides a good diagnostic for the purity of low-energy states? 

We provide a simple answer to this question in \S\ref{sec:epsilonprescription} below, but before doing so we complete the discussion of field redefinitions in interacting theories. Although field redefinitions cannot remove all interactions, they can remove some of them and this makes them useful since they can identify redundant combinations of interactions that do not contribute to observables (including in particular the low-energy sector's purity). 

For example the cubic nonderivative interaction can be removed to leading order in $1/M$ by performing the redefinition  
\be\label{CubicRedef2}
  \phi \to \phi + \frac{a\, \mfg}{M^2}\, \phi \, \sigma + \cO(M^{-3})  
  \quad \hbox{and} \quad
  \sigma \to \sigma  +  \frac{b\,\mfg}{M^2} \, \phi^2 + \cO(M^{-3}) \,.
\ee
which preserves the symmetry $\phi \to - \phi$. This transforms the scalar potential to
\be
   V = \tfrac12 M^2 \phi^2 + \tfrac12 m^2 \sigma^2 + \mfg \phi^2 \sigma 
   \to 
  \tfrac12 M^2 \phi^2 + \tfrac12 m^2 \sigma^2 + \mfg \phi^2 \sigma \left(1 + a + \frac{bm^2}{M^2}
   \right) + \cO(\mfg^2) \,. 
\ee
showing how the $\phi^2 \sigma$ term can be suppressed by $m^2/M^2$ with the choice $a = -1$. 

The interaction does not completely disappear however because the kinetic terms become
\be
   \tfrac12 \Bigl[ (\partial \phi)^2 + (\partial \sigma)^2 \Bigr] 
  \to  \tfrac12 (\partial \phi)^2 \left(1 - \frac{2 \mfg}{M^2} \, \sigma \right)   + \tfrac12 (\partial \sigma)^2   -  \left(1 - 2b  \right)
  \frac{\mfg}{M^2} \, \phi \, (\partial \phi) (\partial \sigma) + \cO(\mfg^2)
\ee
showing that the $\phi \, \partial \phi \, \partial \sigma$ is avoided if we also choose $b = - \frac12 \, a = \frac12$.  With these choices the upshot of such a transformation at order $1/M^2$ is to convert $\mfg \to \frac12 \mfg m^2/M^2$ as the coefficient of the cubic $\mfg \sigma^2 \phi$ interaction and to generate the cubic $ \frac12 \kappa \sigma (\partial \phi)^2$ interaction with $\kappa = - 2\mfg/M^2$. This moves the leading contribution to the purity from the non-derivative to the derivative cubic interaction (but does not remove it). This freedom will become relevant when we discuss questions involving renormalization in the purity calculation.

\subsection{$i\epsilon$ prescription}
\label{sec:epsilonprescription}

Given the above discussion, it might seem as if it is difficult to compute open-system quantities when high-energy states are integrated out. This section argues that this is not true: the correct energy projection required when integrating out heavy states is automatically built in by the $i\epsilon$ prescription associated with the Wightman function. 

We start by clarifying precisely which $i\epsilon$ prescription we have in mind, because related prescriptions are used in a variety of contexts -- particle physics and cosmology, condensed matter, quantum optics... -- for a variety of reasons \cite{Sieberer_2016}. The $i \epsilon$ prescription relevant here is the requirement that the Wightman function, $W(\bfx, t; \bfx', t')$ be evaluated with time differences $t-t_0$ that have a small negative imaginary part. This imaginary part is required to ensure that the sum over states appearing in 
\be
   \langle 0 | \, \phi(x) \, \phi(x') \, | 0 \rangle = \int \exd^3p \; \langle 0 | \phi(0)  |\bfp \rangle \, \langle \bfp | \phi(0)  |0\rangle \, e^{ip \cdot (x-x')} \,,
\ee
converges for large $|\bfp|$, where $p \cdot (x-x') = p_\mu (x-x')^\mu =  - \omega(p) \, (t- t') + \bfp \cdot (\bfx - \bfx')$ and $\omega(p) = \sqrt{\bfp^2 + M^2}$ is the field's dispersion relation. This is similar in spirit to its use in quantum optics, where the $i \epsilon$ prescription is used to regulate the finite size of a detector, with the limit $\epsilon \rightarrow 0$ removing the unresolved physics. In this UV role $\epsilon$ can be regarded as an inverse cutoff $\Lambda^{-1}$ in energy above which details of the theory are lost. This is conceptually different from other uses of $i\epsilon$ that are instead designed to implement initial conditions by selecting the vacuum as the initial state (which is at heart an IR issue).\footnote{Similar reasons are used in particle and condensed-matter physics where the Feynman $i\epsilon$ prescription ensures the initial state within time-ordered correlators is the ground state.}

For the present purposes it is the role of the $i\epsilon$ to act as a UV regulator in the Wightman function that is most important. After all this explicitly suppresses the contributions of energy eigenstates based on their energy eigenvalue, precisely as is required when projecting out high-energy states in the decoupling basis. This can be seen explicitly in the examples of \S\ref{sec:ConcreteExamples}, when these are performed in the `sudden' approximation (rather than with interactions turned on adiabatically). For instance, the purity evolution depends on the real part of \pref{Appinterimintegral2}, and -- as noted in \cite{Burgess:2024eng} -- this is exponentially small in $1/M$ when $M \to \infty$ with $\epsilon = - \hbox{Im}\,(t-t_0)$ is held fixed. 

We see -- see \cite{Burgess:2024eng} for details -- that the limits $\epsilon \to 0$ and $M \to \infty$ do not commute, and this failure to commute is crucial for making decoupling manifest. 
\begin{enumerate}
	\item If one first expands in $1/M$ and then takes $\epsilon \to 0$, the system's state remains pure up to exponentially small corrections, consistent with expectations of \S\ref{sec:PurityTheorem} for integrating out using the exact (decoupling) energy eigenbasis.
	\item If one first takes $\epsilon \to 0$ and then expands in $1/M$, the system's state is found to be mixed by an amount proportional to a power of $\mathcal{O}(1/M)$, consistent with the calculations of \S\ref{sec:ConcreteExamples} in the decohered basis. 
\end{enumerate}

This failure to commute makes sense physically, since $\epsilon \sim 1/\Lambda$ sets the shortest scale in the Wightman function that can be resolved. Heavy physics beyond the reach of EFT methods satisfies $M > \Lambda$ and so provides effects that are too rapid to be detected by low-energy observers. When quantifying the effects of such modes it would be wrong to set detector resolution to zero, as one implicitly does when taking $\epsilon \to 0$ before expanding in powers of $1/M$. The decoupling limit is manifest in this limit because nonzero $\epsilon$ automatically ensures that projecting out the environment is done in a way that discriminates against energy eigenstates without first having to explicitly diagonalize the heavy sector. 

If, on the other hand, $M < \Lambda$ then effects of order $1/M$ can in principle be resolved within low-energy detectors. In this case we can first set $\epsilon \to 0$ and later expand in powers $1/M$.  In this case there can be sizeable effects on the purity as found above when calculating in the decohered basis.\footnote{It is instructive to consider the analogous interpretation in the condensed matter perspective. In the former case where one first expands in $1/M$, keeping the $i \epsilon$ prescription up to the end ensures that the system obtains the desired statistics, that is the one of a vacuum state with zero occupation number. On the contrary, in latter case where one first expands in $\epsilon$, one first places the system in its vacuum configuration, then the environment is able to disrupt it by generating a small occupation number in $\mathcal{O}(1/M)$, which \textit{in fine} generates a sizeable offset of the purity.}

Both computations (decoupled and decohered basis) are right in their own context, and they disagree because they are the answers to different questions. Which is relevant in any particular physical application depends on the relative size of $M$ and $\Lambda$.

\section{UV divergences}
\label{sec:UVdivergences}

We pause in this section to discuss how ultraviolet divergences appear within decoherence calculations, since this is closely related to issues of decoupling. As the previous section argues, UV divergences can be regulated by the Wightman function's $i\epsilon$ prescription, and it is typically true that decoherence calculations remain finite as a UV cutoff goes to infinity with $\epsilon$ held fixed, in agreement with the decoupling limit. In this section we instead ask how UV divergences appear in the opposite limit where $\epsilon$ is taken to zero before identifying the divergences.

In particular, it is tempting to argue that UV divergences must simply cancel within any decoherence calculation simply because UV divergences are usually renormalized into effective couplings in the local effective Lagrangian, order by order in powers of a UV scale $1/\Lambda$. But this means that they can be captured by some sort of Hamiltonian evolution, which we know never changes pure states into mixed ones (so they do not contribute nontrivially to the purity). We argue here why this reasoning is incorrect (see also \cite{Agon:2017oia}). 

The more precise argument against there being divergences in $\partial_t \gamma$ goes as follows. Suppose $H = H_0 + g H_{\rm int}$ where $g$ is the small parameter on which perturbation theory is based. In the usual renormalization program divergences are normally cancelled by making divergent $\cO(g)$ shifts to the couplings in $H$. Recall that the evolution of the reduced density matrix is given by (see eq.~\pref{rhoevoFApp})
\bea \label{rhoevoF}
  \partial_t \varrho(t) &\simeq&  -i g\left[ \ol H_{\rm int}(t), \, \varrho \right] + g^2\sum_{ij} \int_{t_0}^t \exd s \int \exd^3x \int \exd^3y \bigl\{ \left[ \cO^j_{\rm sys}(s,\bfy) \, \varrho_0 \,, \cO^i_{\rm sys}(t,\bfx) \right] \cW^{ij}(t,\bfx; s,\bfy)  \\
  && \qquad\qquad \qquad \qquad \qquad \qquad \qquad\qquad \qquad + \left[ \cO^i_{\rm sys}(t,\bfx) \,, \varrho_0 \cO^j_{\rm sys}(s,\bfy) \right] \cW^{ji} (s,\bfy; t,\bfx) \bigr\}+ \cdots \,,\nn
\eea 
where we write $H_{\rm int} = \sum_i \int \exd^3x \; \cO^i_{\rm sys} \otimes \cO^i_{\rm env}$ and $\cW^{ij}$ denotes the Wightman function for the operators $\cO^i_{\rm env}$. Here $\ol H_{\rm int}$ denotes the average of the interaction Hamiltonian over the environment. Decoherence first arises at second order in $g$ because the first-order term has the form of a Liouville equation and so describes Hamiltonian evolution (that cannot change the purity of an initially pure state). But if divergences in the leading second-order terms in \pref{rhoevoF} are cancelled by $\cO(g)$ corrections to effective couplings in $H_{\rm int}$ then those corrections must appear in the term of  \pref{rhoevoF} that is linear in $H_{\rm int}$ in order to contribute at the same order in $g$. This means they cannot contribute to the decoherence since decoherence receives no contributions at all from \pref{rhoevoF} at first order in $H_{\rm int}$.

This argument cannot be the whole story because any explicit calculation of nonzero purity evolution necessarily depends on some of the coupling constants and {\it all} couplings can be expected to receive divergent renormalizations at some order in perturbation theory, since they must absorb divergences in {\it all} observables and not just the purity evolution. The divergent renormalization of these couplings must therefore appear in the $\cO(g^2)$ term of \pref{rhoevoF}, and something must cancel these divergences if the final purity evolution is to remain well-defined and regularization independent.  There are two places where this cancellation can come from: an explicit divergence in the $\cO(g^2)$ term (suggesting these should in general exist) and potential divergences in the precise definition of the trace that defines the contribution of the environment.

\subsection*{An example revisited}

It is useful to have an explicit example in mind to help sort out what is going on. All of the calculations described in \S\ref{sec:ConcreteExamples} are explicitly UV finite, yet it is also not that hard to come up with examples where such divergences really do appear. A simple example returns to the example of \S\ref{sec:DerivativeCubic} for which the interaction is cubic and involves derivative interactions, but allows the time and spatial derivatives to have different normalization: 
\begin{eqnarray}\label{dim5IntNR}
 \cL_{\rm int} =  \tfrac12  \Bigl[  \kappa_t \,  \dot\phi^2 -   \kappa_s \, \nabla \phi \cdot  \nabla \phi \Bigr] \sigma \,,
\end{eqnarray}
where $\kappa_t$ and $\kappa_s$ are now independent couplings with dimension length. The Lorentz-invariant example of \S\ref{sec:DerivativeCubic} can be obtained as the limit $\kappa_t = \kappa_s = \kappa$. 

The calculation of the purity for the field $\sigma$ goes through much as before, so we record here only those steps that differ from previous examples. Computing the interaction Hamiltonian shows that its leading form corresponds to the choice $\cO_{\rm sys} = \sigma$ but $\cO_{\rm env}=  -\tfrac12 \kappa_t \, \dot \phi^2 + \tfrac12 \kappa_s \, \nabla \phi \cdot \nabla \phi$. Computing the correlator $\cW(x,x')  =  \langle 0 | \cO_{\rm env}(x) \, \cO_{\rm env}(x') | 0 \rangle   - \langle 0 | \, \cO_{\rm env}(x) \, | 0 \rangle    \langle 0 | \, \cO_{\rm env}(x') | 0 \rangle$ leads to ({\it c.f.} eq.~(\ref{cWcubicderiv}))
\begin{equation}
 \cW_{\bmk}(t_1,t_2) = \frac{1}{ (8 \pi )^5 k} \int_{k}^\infty \exd P \int_0^k \exd Q\;  \frac{ (P^2 - Q^2) \Bigl[\kappa_s ( P^2 + Q^2 - 2k^2 ) - 4 \kappa_t \omega_+\omega_-\Bigr]^2}{ \omega_+ \omega_- } \;  e^{ - i \left[ \omega_+ + \omega_- \right] (t_1 - t_2 )} \,,
\end{equation}
which when used in \pref{PurityMasterEqFNested}  gives
\begin{equation} \label{PP3NR}
\gamma_{\bmk} = 1 - \frac{ 1}{  (8\pi)^5 k \omega_\sigma } \int_k^\infty \exd P \int_0^k \exd Q\; \frac{ (P^2 - Q^2) \Bigl[\kappa_s ( P^2 + Q^2 - 2k^2) - 4 \kappa_t \omega_+ \omega_- \Bigr]^2 }{ \omega_+ \omega_- \big( \omega_\sigma + \omega_+ + \omega_- \big)^2 } \,,
\end{equation}
where $\omega_\pm$ are as defined below eq.~\pref{cWcubicderiv}. The large-$P$ limit of the integrand now is
\bea
 && \frac{ (P^2 - Q^2) \Bigl[ \kappa_s (P^2 + Q^2 - 2k^2) - 4 \kappa_t \omega_+ \omega_- \Bigr]^2}{ \omega_+ \omega_- \big( \omega_\sigma + \omega_+ + \omega_- \big)^2 } \simeq 4(\kappa_s - \kappa_t)^2 P^2 - 8 (\kappa_s - \kappa_t)^2 \omega_\sigma P \\
  && \qquad\qquad\qquad\qquad +4(\kappa_t-\kappa_s) \Bigl[ 4 \kappa_s(8M^2 -2Q^2-3\omega_\sigma^2 +4 k^2)+ \kappa_t(3\omega_\sigma^2 - 2 Q^2)   \Bigr] \nn\\
 && \qquad\qquad\qquad\qquad\qquad + \frac{16 \omega_\sigma}{P}(\kappa_t - \kappa_s) \Bigl[ \kappa_s(Q^2 + \omega_\sigma^2 - 5M^2 - 2k^2) + \kappa_t (M^2 + Q^2 - \omega_\sigma^2) \Bigr]  + \cO(P^{-2})  \nn
\eea
and so the result is only UV finite only in the special case $\kappa_s = \kappa_t$. Regulating using a momentum cutoff $P<\Lambda$ leads to the divergent contribution
\bea \label{PPFNR}
\gamma_{\bmk}^\infty &\simeq& -\frac{ 4(\kappa_t - \kappa_s)}{  (8\pi)^5  \omega_\sigma } \left\{ \tfrac13 (\kappa_t - \kappa_s) \Lambda^3 - (\kappa_t - \kappa_s) \omega_\sigma \Lambda^2   +  \left[ 4 \kappa_s \left(8M^2 -3\omega_\sigma^2 +\tfrac{10}3 k^2 \right) - \kappa_t \left(3\omega_\sigma^2 - \tfrac23 k^2 \right)  \right] \Lambda \phantom{\frac12} \right.\nn\\
 && \qquad\qquad\qquad\qquad\qquad\left.  -4 \omega_\sigma  \Bigl[ \kappa_s(  \omega_\sigma^2 - 5M^2 - \tfrac53 k^2) + \kappa_t (M^2 + \tfrac13 k^2 - \omega_\sigma^2) \Bigr]   \log\left(\frac{\Lambda}{\mu} \right)  \right\}  \,,
\eea
which indeed vanishes in the limit considered earlier: $\kappa_s = \kappa_t = \kappa$.

\subsubsection*{Renormalization}

What cancels these divergences when $\kappa_s \neq \kappa_t$? To find out we must include all interactions that can arise to the order we work. In the present instance we want the most general form possible for $\cL = \cL_0 + \cL_{\rm int}$ consistent with the $\phi \to - \phi$ symmetry, time-reversal and rotation invariance where we care only about interactions that couple $\sigma$ to $\phi$ out to dimension five. These criteria give
\be \label{FreeL0}
\cL_0 =  \tfrac12 \Bigl[ \dot \sigma^2 +  \dot \phi^2 - c^2_\sigma (\nabla \sigma)^2 - c^2_\phi \, (\nabla \phi)^2  -  m^2 \sigma^2 -  M^2 \phi^2 \Bigr] \,,
\ee
where $c_\phi$ and $c_\sigma$ denote the `sound speed' for each field (with $1 - c^2$ regarded as being perturbatively small) and the interactions can be classified by dimension: $\cL_{\rm int} =  \cL^{(3)}_{\rm int} + \cL^{(4)}_{\rm int} + \cL^{(5)}_{\rm int} + \cdots$, with
\bea \label{CompleteSet}
    &&\qquad\qquad\qquad  \cL^{(3)}_{\rm int} = - \mfg  \sigma \phi^2 \,, \qquad
    \cL^{(4)}_{\rm int} =   - \tfrac14 \lambda_{22} \sigma^2 \phi^2  \\
    &&  \cL^{(5)}_{\rm int} = - \lambda_{14} \sigma \phi^4 - \lambda_{32} \sigma^3 \phi^2 + \tfrac12 \kappa_t \sigma \dot \phi^2 - \tfrac12 \kappa_s \sigma (\nabla \phi)^2 + \hat \kappa_t \phi \, \dot \sigma \dot \phi  - \hat \kappa_s \phi \, \nabla \sigma \cdot \nabla \phi \,, \nn
\eea
and so on. 

Not all of these interactions are independent of one another, as was seen in \S\ref{ssec:DecovsDeco} which showed how the cubic $\mfg$ interaction could be removed using a nonlinear field redefinition, at the expense of changing the coefficients $\kappa_i$ and $\hat \kappa_i$. It is useful to use this freedom to identify a minimal subset of couplings, $\cO_i$. The point is that second-order contributions of the complete set of such minimial interactions can appear in the purity evolution and there are two potential sources for counterterm contributions that can absorb the UV divergences that arise. One of these is in a `wavefunction' renormalizations that mix the operators $\cO_i \to \cZ_{ij} \cO_j$. Another is in the possible necessity of performing additional nonlinear field redefinitions
\be\label{CubicRedef22}
  \phi \to Z_1^{1/2} \, \phi + \alpha\, \phi \, \sigma + \cdots   
  \quad \hbox{and} \quad
  \sigma \to Z_2^{1/2} \, \sigma  +  \beta \, \phi^2 + \cdots  \,,
\ee
in order to return the most general couplings into the minimial set $\cO_i$. Divergences in $\cZ_{ij}$ and in coefficients like $\alpha$ or $Z_i$ can cancel the explicit divergences appearing in the calculation of purity. It is important that the presence of nonzero masses allows operators of different dimensions to mix in this way, and so for lower-dimensional couplings to cancel UV divergences obtained using the higher-dimension couplings.

In the case of interest here the divergences found above using the interaction \pref{dim5IntNR} can in principle cancel against divergences coming from the contribution to decoherence produced by the other interactions in \pref{CompleteSet}, or with divergent parameters in transformations like \pref{CubicRedef22}. For example performing \pref{CubicRedef22} on \pref{FreeL0} generates the counterterms
\bea \label{FreeCT}
\cL_0 &=&  \tfrac12 (Z_1-1) \Bigl[ \dot \sigma^2  - c^2_\sigma (\nabla \sigma)^2 -  m^2 \sigma^2 \Bigr] + (Z_2-1) \Bigl[\dot \phi^2 - c^2_\phi \, (\nabla \phi)^2   -  M^2 \phi^2 \Bigr]  \\
&&\qquad\qquad + \alpha \, \sigma \Bigl[ \dot \phi^2 - c_\phi^2 (\nabla \phi)^2 \Bigr] + (\alpha + 2 \beta) \phi \, \dot \sigma \, \dot \phi - (\alpha  c_\phi^2 + 2\beta c_\sigma^2) \, \phi \, \nabla \phi \cdot \nabla \sigma + \cdots  \,, \nn
\eea
and so in particular shifts the bare couplings by
\be \label{alphadepct}
  \delta \mfg = \alpha M^2 + \beta m^2 \,, \quad
   \delta \kappa_t = 2 \alpha \,, \quad \delta \kappa_s = 2 \alpha \, c_\phi^2 \,, \quad
   \delta \hat \kappa_t = \alpha + 2 \beta \,,\quad 
   \delta \hat \kappa_s = \alpha c_\phi^2 + 2 \beta c_\sigma^2 \,,
\ee
and so on. In principle one computes the Wightman function including all of these interactions and counterterms and uses the counterterms to cancel the explicit divergences.

For instance, suppose the breaking of Lorentz invariance at tree level in the Lagrangian is controlled by a small parameter, so $0 < |\kappa_t - \kappa_s| \ll \bar \kappa := \frac12(\kappa_s + \kappa_t)$. At leading order in this small parameter $c_\phi = c_\sigma = 1$ and $\hat \kappa_s = \hat \kappa_t = 0$ (as was implicitly chosen when evaluating the integrals appearing in the purity above) and so the counterterm $\alpha$ in \pref{alphadepct} appears in the combination $\delta \kappa_t = \delta \kappa_s = \delta \kappa = 2\alpha$. Inspection of \pref{PPF} shows that perturbing the finite result from \S\ref{sec:DerivativeCubic} by this amount gives 

\begin{equation} \label{PPF2}
\gamma_{\bmk}^{\rm ct} \simeq  - \frac{ \bar \kappa \delta \kappa  M^3 }{ 512 \pi^4 \omega_{\sigma} }  =  - \frac{ \bar \kappa \alpha  M^3 }{ 256 \pi^4 \omega_{\sigma} }  \,,
\end{equation}
so $\alpha$ can be chosen to cancel the divergence in the $\omega_\sigma$- and $k$-independent UV divergence displayed in \pref{PPFNR} (repeated here for convenience, linearized about $\bar \kappa$)
\be \label{PPFNR4}
\gamma_{\bmk}^\infty \simeq -\frac{ 4(\kappa_t - \kappa_s)}{  (8\pi)^5  \omega_\sigma } \Bigl[  32\bar \kappa M^2  \Lambda+ \cO(\omega_\sigma, k, \kappa_t - \kappa_s) \Bigr]  \,,
\ee
by choosing 
\be
   \alpha = - \frac{\Lambda}{\pi M} (\kappa_t - \kappa_s) \,.
\ee

To summarize, earlier sections gave general arguments that decoupling prevents heavy fields from contributing to the purity evolution of low-energy states. It is tempting to conclude from this that calculations of purity evolution must simply be UV finite (as has indeed been found to be true for specific examples). The example considered in this section shows that this tempting conclusion is wrong: UV divergences can arise in purity calculations. The presence of divergences also seems reasonable given that all effective couplings generically contain divergent renormalizations (counterterms), as required to renormalize divergences in other types of observables. Explicit divergences in decoherence calculations are precisely what is required for cancel these counterterm divergences, and divergences to do with the mixing of composite operators under renormalization (see for example \cite{Collins:1984xc}). Although the above calculations are suggestive and we regard it as plausible that divergent decoherence calculations can be renormalized in the same way as are divergences in other observables, proving this remains an open question that goes beyond the scope of this paper. 

\section{Conclusions}
\label{sec:Conclusions}

Effective field theories and open quantum systems are similar in that they are rooted in practicality: the former acknowledges that no measurement can achieve perfect accuracy, while the latter recognizes that no system is ever completely isolated.  Both frameworks aim to reduce the degrees of freedom required to accurately describe a given physical situation, with EFTs specifically designed to efficiently balance retained and discarded information as early as possible in any calculation. An apparent tension arises between the expected decoherence of quantum systems interacting with their environment \cite{Zurek:1981xq, Zurek:1982ii, Joos:1984uk} and the equally expected autonomous evolution of low-energy degrees of freedom \cite{Weinberg:1978kz, EFTBook}. However, rather than representing conflicting principles, we here argue that these two perspectives offer complementary insights into the reductionism that enables us to describe physical systems with a reduced set of degrees of freedom.

In this article, we have illustrated this tension by showing through various examples in Sec.~\ref{sec:ConcreteExamples} that tracing over the Hilbert space of heavy field results in finite $1/M$ corrections to the system's purity. This indeed contrasts with the expectation set by the decoupling theorem discussed in Sec.~\ref{sec:PurityTheorem}. One might then ask whether a low-energy EFT can be constructed in which the state remains pure. In Sec.~\ref{sec:Decoupling}, we demonstrate that it is indeed possible to identify a bipartition where the purity is systematically more than polynomially suppressed in the expansion parameter. This approach relies on the resolution scale $\epsilon \sim 1/\Lambda$ being much larger than the heavy mass corrections, requiring the expansion in the heavy mass to be performed first, followed by the standard $i \epsilon$ prescription. Physically, when $M \ll \Lambda$, the heavy field can deviate from its vacuum state, resulting in finite corrections to the purity. Conversely, when $\Lambda \ll M$, the heavy field remains fixed in the vacuum, ensuring the preservation of purity.
Identifying a bipartition consistent with decoupling expectations might suggest the tempting conclusion that the purity should be UV finite. However, as we show in Sec.~\ref{sec:UVdivergences}, this is not the case due to the dependence of purity on the effective couplings of the theory.

While our results provide a new perspective on the relationship between decoherence and decoupling, they also raise important questions about the use of quantum information measures (such as purity) in EFTs --- highlighting the need for further exploration in this area. Although decoupling is well-established in particle physics, its validity when considering the entanglement structure of quantum field theories in both flat and curved space-time remains an open question, particularly when it is the explicit time evolution of states and observables that are of interest. As our calculations suggest, decoupling persists as usual and no tension arises if one adiabatically turns the interactions on and off at past and future infinity --- the in-out standard setup for the $S$-matrix, where effective field theories (EFTs) are known to perform exceptionally well. It is the departure from this conventional setup, when exploring in-in (or Schwinger-Keldysh) time evolution, that introduces interesting complications and potential insights into the possible extensions of the EFT paradigm (see also \cite{Salcedo:2022aal}).

A key aspect of our work lies in the selection of degrees of freedom used in the in-in (Schwinger-Keldysh) contour, implicitly employed for computing the state's purity. This is well-known in condensed matter and cosmology where different degrees of freedom correspond to inequivalent vacua, ultimately leading to distinct bipartitions. In this context, the choice of canonical variables conveys physical information about the system, unlike the familiar reparametrization invariance of the $S$-matrix. Specifically, redefining the degrees of freedom alters the amount of information they share, as demonstrated in \cite{Salcedo:2022aal, Sou:2022nsd, Ning:2023ybc, Sou:2024tjv}, where boundary terms and total derivatives modify the purity explicilty. 

Given the dependence of entropy measures such as purity on the choice of field variables, clear guidelines may be needed to define system/environment bipartitions. Such guidance could come from experimental considerations or from the presence of symmetries and conserved charges. As a byproduct of our analysis, it would be interesting to explore whether purity and entanglement entropy could be used to identify decoherence-free subsectors that evolve autonomously. Such subsectors have been identified through explicit computations, as seen in \cite{Gundhi:2024fqv} for flat space and \cite{Colas:2022kfu} for cosmological inflation. In the latter case, the background evolution (further examined in \cite{Colas:2024xjy}) plays a key role in suppressing derivative interactions governed by gradients. Their decay at late times on super-Hubble scales allows for the emergence of a distinctly non-Markovian recoherence phenomenon \cite{Colas:2022kfu}. From the present perspective the recoherence example can be understood as a field that undergoes decoherence at early times and then effectively decouples at late times (tending exponentially close to a pure state) due to a time-dependent system-environment coupling that shuts off at late times. The time-dependence of the couplings effectively realign the decohered and decoupled bases. Gaining a deeper understanding of the prevalence of this process for other types of processes in cosmology could prove extremely valuable.

Our findings may provide a foundation for future studies exploring divergences \cite{Agon:2017oia}, renormalization \cite{Baidya:2017eho}, and power counting \cite{Burgess:2009ea} in Open EFTs. They could also pave the way toward a unified approach to the $i \epsilon$ prescription in cosmology \cite{Adshead:2009cb, Kaya:2018jdo, Albayrak:2023hie}, condensed matter \cite{Sieberer_2016}, and quantum optics \cite{Tokieda:2024soc}. Lastly, the study of decoherence in cosmology and near black holes is still in its early stages. Investigating the role of horizons \cite{Danielson:2022tdw, Danielson:2022sga, Wilson-Gerow:2024ljx, Biggs:2024dgp, Danielson:2024yru} in light of our results offers a promising direction for applying entropy measures \cite{Cheung:2018cwt, Cheung:2019cwi, Dvali:2020wqi, Creminelli:2022onn, Cheung:2023hkq, Aoude:2024xpx, DuasoPueyo:2024rsa,Ueda:2024cyf,Low:2024hvn} in perturbative QFT in curved spacetime.

\section*{Acknowledgements}

We thank Santiago Agui Salcedo, Sebasti\'an C\'espedes, Jerome Martin and Enrico Pajer for many useful discussions. We extend our special thanks to Vincent Vennin for his early collaboration on this work and for reviewing this draft and providing many insightful comments. CB's research was partially supported by funds from the Natural Sciences and Engineering Research Council (NSERC) of Canada. Research at the Perimeter Institute is supported in part by the Government of Canada through NSERC and by the Province of Ontario through MRI. This work has been supported by STFC consolidated grant ST/X001113/1, ST/T000791/1, ST/T000694/1 and ST/X000664/1.

\changelocaltocdepth{1}

\begin{appendix}

\section{Computing purity evolution in interaction picture}
\label{App:InteractionPicture}

This appendix recalls the expressions for purity evolution that are used by the examples in the main text. This section does so using perturbation theory in the interaction picture. To this end we imagine the Hamiltonian breaks up into $H = H_{\rm sys} + H_{\rm env} + H_{\rm int}$ and we perturb in powers of $H_{\rm int}$.  We further assume $H_{\rm int}$ can be broken into a basis of product operators, so that its interaction-picture version has the form
\be
  H_{\rm int} (t) = \sum_i \cO^i_{\rm sys}(t) \, \cO^i_{\rm env}(t) \,.
\ee

\subsection{General formalism}
\label{Appsec:GenForm}

In the interaction picture the full density matrix $\rho(t)$ for the scalar system satisfies the Liouville equation
\be \label{Liouville}
   \partial_t \rho = -i \Bigl[ H_{\rm int}(t) \,, \rho \Bigr] \,.
\ee
So far as measurements of the system are concerned all we really require is the evolution of the reduced density matrix for $\sigma$, defined by tracing out the $\phi$ sector:
\be
   \varrho(t) := \Tr_\phi [ \rho(t) ] \,. 
\ee
In perturbation theory this satisfies the evolution equation
\bea \label{rhoevoFApp}
  \partial_t \varrho(t) &\simeq&  -i \left[ \ol H_{\rm int}(t), \, \varrho \right] + \sum_{ij} \int_{t_0}^t \exd s \int \exd^3x \int \exd^3y \bigl\{ \left[ \cO^j_{\rm sys}(s,\bfy) \, \varrho_0 \,, \cO^i_{\rm sys}(t,\bfx) \right] \cW^{ij}(t,\bfx; s,\bfy)  \\
  && \qquad\qquad \qquad \qquad \qquad \qquad \qquad\qquad \qquad + \left[ \cO^i_{\rm sys}(t,\bfx) \,, \varrho_0 \cO^j_{\rm sys}(s,\bfy) \right] \cW^{ji} (s,\bfy; t,\bfx) \bigr\}+ \cdots \,,\nn
\eea 
where 
\be
   \ol H_{\rm int} = \sum_i \llangle \, \cO^i_{\rm env} \, \rrangle \, \cO^i_{\rm sys}
\ee
where $\llangle \,\cdots\, \rrangle$ is defined below \pref{LowEnergyStates} and 
\be
\label{eq:Wightmann:def}
  \cW^{ij}(t,\bfx; s,\bfy)  := \llangle \, \cO^i_{\rm env}(t,\bfx) \, \cO^j_{\rm env}(s,\bfy) \,\rrangle - \llangle \, \cO^i_{\rm env}(t,\bfx) \, \rrangle \, \llangle \, \cO^j_{\rm env}(s,\bfy) \, \rrangle    \,. 
\ee
These expressions assume the system and environment are uncorrelated at $t = t_0$.

Our diagnostic for decoherence is the purity, $\gamma(t)$, defined as
\be 
   \gamma(t) := \Tr_\sigma [ \varrho^2(t) ]
\ee
and so $0 \leq \gamma \leq 1$, with $\gamma=1$ if and only if the reduced state $\varrho$ is pure. Its rate of change is\footnote{This result assumes $\Tr_\sigma[\varrho_0 \cO^i_{\rm env}(t,\bfx) \varrho_0 \cO^j_{\rm env}(s,\bfy)] = 0$ as would be the case if $\varrho_0 = | \Psi \rangle \, \langle \Psi |$ and $\langle \Psi | \cO^i_{\rm env}(t,\bfx) | \Psi \rangle = 0$, and so is true in particular if $|\Psi \rangle = | \hbox{vac} \rangle$. Even if not initially true it can often be made to be true by appropriately redefining the operators $\cO^i_{\rm env}$.}
\be \label{PurityChangeMasterEqApp}
  \partial_t \gamma =2 \, \Tr_\sigma \left( \varrho \, \partial_t \varrho \right)  =  - 4\sum_{ij} \int_{t_0}^t \exd s \int \exd^3x \int \exd^3y \; \hbox{Re}\Bigl[ W^{ij}(t,\bfx; s,\bfy) \, \cW^{ij}(t,\bfx; s,\bfy) \Bigr] \,,
\ee
where $\cW^{ij}$ is given in \pref{eq:Wightmann:def} and $W^{ij}$ is defined by 
\be \label{Wsysdef}
  W^{ij}(t,\bfx;s,\bfy)  :=  \langle \cO^i_{\rm sys}(t,\bfx) \, \cO^j_{\rm sys}(s,\bfy) \rangle  -  \langle \cO^i_{\rm sys}(t,\bfx) \rangle \, \langle \cO^j_{\rm sys}(s,\bfy) \rangle \,,
\ee
with the expectation value  $\langle \,\cdots\, \rangle = \trA[(\cdots) \,\varrho_{0}]$ taken in the observed system's prepared state. These correlation functions are evaluated by inserting a complete set of states between the two operators that involves a sum over quantum numbers that diverges in the UV. This is dealt with by demanding the time arguments to have small negative imaginary parts so that $e^{-i \omega (t-s)}$ is exponentially damped for large $\omega$.  
 
For translation-invariant systems it is often more useful to use the Fourier representation
\be \label{cWtocWk}
  \cW^{ij}(t,\bfx; s,\bfy)   =   \int \frac{ \exd^3k}{(2\pi)^3} \; \cW^{ij}_{\bf{k}}(t,s) \, e^{ i \bf{k} \cdot (\bf{x} - \bf{y})}
\ee
and the equivalent expression for $W^{ij}(t,\bfx; s,\bfy)$ in terms of $W^{ij}_{\bf{k}}(t,s)$,. With these \pref{PurityChangeMasterEqApp} can be written
\be \label{PurityChangeMasterEqFkApp}
  \partial_t \gamma   =  - 4 \cV \int_{t_0}^t \exd s\int \frac{\exd^3k}{(2\pi)^3} \; \hbox{Re}\Bigl[ W^{ij}_\bmk(t,s) \, \cW^{ij}_{-\bmk}(t,s) \Bigr] \,.
\ee

In the special case where the $\cO^i_{\rm sys}$ are linear in $\sigma$ and when the initial state is chosen so that each mode $\bmk$ is uncorrelated with the others then it is convenient to factorize the subsequent evolution, 
\be
   \varrho(t) = \bigotimes_{\bmk} \varrho_\bmk(t) \qquad \hbox{and} \qquad  \varrho^2(t) =   \bigotimes_{\bmk} \;  \varrho^2_\bmk(t) 
\ee
where we switch to discretely normalized momentum states. In this case the purity of each mode evolves separately and so defining $\gamma_\bmk(t) = \Tr_\sigma [ \varrho^2_\bmk(t) ]$ one finds
\be \label{PurityChangeMasterEqFSumApp}
  \partial_t \gamma_\bmk =- 4 \sum_{ij} \int_{t_0}^t \exd s  \; \hbox{Re}\Bigl[ W^{ij}_\bmk(t,s) \, \cW^{ij}_{-\bmk}(t,s) \Bigr] \,.
\ee
This integrates to give a useful expression for the purity as a function of time
\be \label{AppPurityMasterEqFNested}
   \gamma_\bmk(t)       =  \gamma_\bmk(t_0) - 2 \sum_{ij} \int_{t_0}^t \exd s' \int_{t_0}^t \exd s  \; \hbox{Re}\Bigl[ W^{ij}_\bmk(s',s) \, \cW^{ji}_{-\bmk}(s',s) \Bigr] \,,
\ee 
which uses
\be \label{conjugationproperty}
   [W^{ij}_\bmk(t,s)]^* = W^{ji}_{-\bmk}(s,t)  \qquad \hbox{and} \qquad [\cW^{ij}_\bmk(t,s)]^* = \cW^{ji}_{-\bmk}(s,t)  \, ,
\ee
when $\cO^i_{\rm sys}$ and $\cO^j_{\rm env}$ are hermitian. There is a subtlety in evaluating the real part in these expressions because the time arguments of the Wightman functions themselves carry a small negative imaginary part.

There are two natural choices when performing the time integral. Doing so with the system and environment in their ground states at finite $t_0$ and using time-independent couplings corresponds to turning on the interaction $\cL_{\rm int}$ suddenly at $t=t_0$. In this case one finds integrals of the generic form
\bea \label{Appinterimintegral2}
  g \int_{t_0}^t \exd s \; e^{-i(\omega_\sigma + \omega_\phi)s} 
  &=& \frac{ig}{\omega_\sigma + \omega_\phi} \Bigl[ e^{-i (\omega_\sigma + \omega_\phi)t} -  e^{-i (\omega_\sigma + \omega_\phi)t_0} \Bigr] \nn\\
  &=&  \frac{2 g}{\omega_\sigma + \omega_\phi} \sin\Bigl[ \tfrac12 \, (\omega_\sigma + \omega_\phi)(t-t_0) \Bigr] \, e^{-\frac12 i(\omega_\sigma + \omega_\phi)(t+t_0)} \,,
\eea
where $g$ is the relevant coupling constant appearing in $\cL_{\rm int}$.

Alternatively one can instead imagine sending $t_0 \to - \infty$ and turning on $\cL_{\rm int}$ by regarding the couplings $g(t)$ as being time-dependent, but varying only adiabatically slowly. In this case the integral \pref{Appinterimintegral2} is replaced by
\bea \label{AppinterimintegralAdia2}
  \int_{-\infty}^t \exd s \; g(s) \, e^{-i(\omega_\sigma + \omega_\phi)s} &=& \left[ \frac{i g(s) \, e^{-i(\omega_\sigma + \omega_\phi)s}}{\omega_\sigma + \omega_\phi} \right]^t_{-\infty} -\frac{i}{\omega_\sigma + \omega_\phi} \int_{-\infty}^t \exd s \; \partial_s g(s) \, e^{-i(\omega_\sigma + \omega_\phi)s} \nn\\
  &=& \frac{ig(t)}{\omega_\sigma + \omega_\phi} \; e^{-i (\omega_\sigma + \omega_\phi)t} -\frac{i}{\omega_\sigma + \omega_\phi} \int_{-\infty}^t \exd s \; \partial_s g(s) \, e^{-i(\omega_\sigma + \omega_\phi)s}    \,,
\eea
where the first equality integrates by parts and the second equality uses $g(-\infty) = 0$. The adiabatic approximation assumes $g$ changes with time so slowly that the integration over $\partial_s g(s)$ can be neglected. The adiabatic result is equivalent to what would have been obtained if we had kept $g$ constant and had taken the lower limit of \pref{Appinterimintegral2} as $t_0 \to -\infty (1 + i \epsilon)$ with positive $\epsilon \to 0$ at the end of the calculation:
\be \label{Appinterimintegral22}
 g \lim_{\epsilon \to 0} \int_{-\infty(1+i\epsilon)}^t \exd s \; e^{-i(\omega_\sigma + \omega_\phi)s} 
  =  \frac{ig}{\omega_\sigma + \omega_\phi}  \; \; e^{-i(\omega_\sigma + \omega_\phi)t}  \,,
\ee
where $g$ is the late-time value to which $g(t)$ asymptotes. Taking the imaginary part of the time difference to infinity in the remote past projects out the nonadiabatic contribution. Notice that the adiabatic result in general does not agree with the time average of the sudden result because the sudden result also includes nonnegligible contributions from the $\partial_s g$ term in \pref{AppinterimintegralAdia2}.

\subsection{Examples}
\label{Appsec:ConcreteExamples}

This section applies the above formalism to the concrete purity calculations described in the main text.  These examples all use a Lagrangian of the form
\be
\cL = -  \left[ \frac12 (\partial \phi)^2 + \frac12 (\partial \sigma)^2 + \frac12 \, M^2 \phi^2 + \frac12 m^2 \sigma^2   \right] + \cL_\rint\,,
\ee
with different choices for the interaction term $\cL_\rint$. 

\subsubsection{Mass mixing}
\label{Appssec:MassMixing}

We first consider mass mixing, for which
\be
   \cL_\rint = -    \mu^2 \, \phi \, \sigma \,.
\ee
This system can be solved exactly by diagonalizing the mass matrix. The two mass eigenstates in this case are $\phi_\pm$, given by
\be  \label{Appmassmixingeigenbasis}
   \left( \begin{matrix} \phi \\ \sigma \end{matrix} \right) = \left( \begin{matrix} \cos\vartheta & -\sin\vartheta \\  \sin \vartheta & \cos \vartheta \end{matrix} \right) \left( \begin{matrix} \phi_+\\ \phi_- \end{matrix} \right)  
   \quad \hbox{with} \quad
   \tan 2\vartheta = \frac{2\mu^2}{M ^2 - m ^2} \,,
\ee
and have squared masses 
\be\label{Appmasseigenvalues}
   M^2_\pm = \frac12 \Bigl[ (m ^2+M ^2) \pm \sqrt{(m ^2-M ^2)^2 + 4\mu^4} \Bigr]  \,.
\ee
These are both positive when $0 \leq \mu^2 \leq M  m $ (as we henceforth assume). 

The perturbative limit in this case assumes $\mu^2 \ll M^2 - m^2$, in which case
\be\label{Appmasseigenvalues+-}
   M^2_+  \simeq M^2 + \frac{ \mu^4}{M^2 - m^2}    \quad \hbox{and} \quad
   M^2_-  \simeq m^2 -  \frac{ \mu^4}{M^2 - m^2}   \,, 
\ee
while the redefinition \pref{Appmassmixingeigenbasis} becomes 
\be\label{AppDiagonalizationRedef}
  \phi \simeq \phi_+ - \frac{\mu^2 \phi_- }{M^2-m^2}  
  \quad \hbox{and} \quad
  \sigma \simeq \phi_- + \frac{\mu^2 \phi_+}{M^2-m^2}      \,.
\ee 
To leading order in $\mu^2$ the purity is given by \pref{AppPurityMasterEqFNested}, whose evaluation requires the Wightman functions for the operators $\cO_{\rm sys} = \sigma$ and $\cO_{\rm env} = \mu^2 \phi$. Using the flat-space vacuum as the initial state then implies that these Wightman functions evaluate in momentum space to
\be \label{AppFlatWk}
 W_{\bmk}(t,s) =  \frac{1}{2 \omega_\sigma(k)} \, e^{-i\omega_\sigma(k)(t-s)} \quad \hbox{and} \quad
 \cW_{\bmk}(t,s) =  \frac{\mu^4}{2 \omega_\phi(k)} \, e^{-i\omega_\phi(k)(t-s)} \,,
\ee
where $\omega_\phi(k) = \sqrt{k^2 + M^2}$ and $\omega_\sigma(k) = \sqrt{k^2 + m^2}$ are the appropriate dispersion relation for each field. 

Performing the time integrals in the sudden approximation one finds in this way the following expression for the purity of the $\sigma$ state given that both $\sigma$ and $\phi$ are initially prepared in their vacuum at $t = t_0$ \cite{Burgess:2024eng}:  
\be  \label{AppPurityMasterEqFflat}
  \gamma_\bmk^s(t)   =  1  -  \frac{2\mu^4}{\omega_\sigma \omega_\phi(\omega_\sigma +\omega_\phi)^2} \, \sin^2 \Bigl[\tfrac12 (\omega_\sigma +\omega_\phi)(t-t_0) \Bigr]  \qquad \hbox{(sudden approximation)}\,.
\ee
Turning the mixing on adiabatically instead gives the time-independent result
\be \label{AppPurityMasterEqFflatAdiabatic}
  \gamma_\bmk^a(t)  \simeq  1  -  \frac{\mu^4}{2\omega_\sigma \omega_\phi(\omega_\sigma +\omega_\phi)^2} \qquad \hbox{(adiabatic approximation)}\,.
\ee
Notice the adiabatic approximation is half as large as is the result obtained by time-averaging the sudden approximation over many periods.

The limiting case when $M \gg k, m, \mu$ is the main focus of the discussion in the main text, and for the above expressions gives the limiting large-$M$ forms
\be  \label{AppPurityMasterEqFflatM}
  \gamma_\bmk^s(t)      \simeq  1  -  \frac{2\mu^4}{\omega_\sigma M^3} \, \sin^2 \Bigl[\tfrac12 M (t-t_0) \Bigr]  \qquad \hbox{(sudden approximation)} \,,
\ee
and
\be \label{AppPurityMasterEqFflatsecular}
  \bar \gamma_\bmk^a(t)  \simeq  1  -  \frac{\mu^4}{2\omega_\sigma M^3} \qquad \hbox{(adiabatic approximation)} \,.
\ee

\subsubsection{Kinetic mixing}
\label{Appssec:KineticMixing}

Next consider  off-diagonal kinetic mixing between $\phi$ and $\sigma$ where 
\be
   \cL_\rint = -   \xi \, \partial_\mu \phi \, \partial^\mu \sigma \,,
\ee
with $|\xi| < 1$. The kinetic terms are returned to canonical form and the mass matrix is rediagonalized through the transformation 
\be  \label{Appkinmixingeigenbasis}
   \left( \begin{matrix} \phi \\ \sigma \end{matrix} \right) = \frac{1}{\sqrt2} \left( \begin{matrix} 1 & 1 \\  -1 & 1 \end{matrix} \right) \left[ \begin{matrix} (1-\xi)^{-1/2} & 0 \\ 0 &  (1+\xi)^{-1/2} \end{matrix} \right] \left( \begin{matrix} \cos\vartheta & -\sin\vartheta \\  \sin \vartheta & \cos \vartheta \end{matrix} \right) \left( \begin{matrix} \phi_+\\ \phi_- \end{matrix} \right)     \,,
\ee
where
\be  \label{Appmassmixingeigenbasiskin}
   \tan 2\vartheta = \left( \frac{M ^2 - m ^2}{M ^2 + m ^2} \right) \frac{\sqrt{1-\xi^2}}{\xi} \,.
\ee
Notice $\xi \to 0$ implies $\vartheta \to \frac{\pi}{4}$ in which case \pref{Appkinmixingeigenbasis} reduces to the trivial transformation with $\phi \to \phi_+$ and $\sigma \to \phi_-$. The resulting mass eigenvalues are
\be\label{Appmasseigenvalueskin}
   M^2_\pm = \frac{1}{2(1-\xi^2)} \Bigl[ (m ^2+M ^2) \pm \sqrt{ (M^2 - m^2)^2 + 4 \xi^2 M^2 m^2} \Bigr]  \,.
\ee

The perturbative limit in this case assumes $2|\xi| Mm \ll M^2 - m^2$, for which dropping all terms involving $\xi^4$ and higher gives
\be\label{Appkinmasseigenvalues+-}
   M^2_+  \simeq M^2 \left[ 1  + \frac{ \xi^2 ( M^2 + m^2)}{M^2 - m^2}  \right]  + \cO(\xi^4)  
   \quad \hbox{and} \quad
   M^2_-  \simeq m^2 \left[ 1  - \frac{ \xi^2 ( M^2 + m^2)}{M^2 - m^2}  \right]   + \cO(\xi^4)   \,,
\ee
while \pref{Appmassmixingeigenbasiskin} becomes
\be
   \vartheta \simeq \frac{\pi}4  - \frac{\xi (M^2+m^2)}{2(M^2-m^2)} + \cO(\xi^2)  
\ee
and so the redefinition \pref{Appkinmixingeigenbasis} is
\be\label{AppDiagonalizationRedef2}
  \phi \simeq \phi_+ + \frac{\xi  m^2 \phi_-}{M^2 - m^2}   + \cO(\xi^2)  
  \quad \hbox{and} \quad
  \sigma \simeq \phi_- - \frac{\xi  M^2 \phi_+}{M^2 - m^2}  + \cO(\xi^2)   \,.
\ee

The presence of time derivatives in $\cL_{\rm int}$ complicates the calculation of purity for the field $\sigma$ because we first must construct the canonical interaction Hamiltonian. Defining the canonical momenta by
\be
  \Pi_\sigma = \frac{\delta S}{\delta \dot \sigma} =   \dot \sigma + \xi \dot \phi  \quad \hbox{and} \quad
  \Pi_\phi = \frac{\delta S}{\delta \dot \phi} =   \dot \phi + \xi \dot \sigma  \,,
\ee 
implies the inverse relation for the velocities is
\be
   \dot\phi = \frac{1}{1-\xi^2} \Bigl( \Pi_\phi - \xi \,\Pi_\sigma \Bigr) \quad \hbox{and} \quad
   \dot \sigma = \frac{1}{1-\xi^2} \Bigl( \Pi_\sigma - \xi \, \Pi_\phi \Bigr) \,.
\ee
The Hamiltonian then becomes $\cH = \Pi_\phi \, \dot \phi + \Pi_\sigma \dot \sigma - \cL = \cH_0 + \cH_{\rm int}$ where  
\be\label{AppH0canon}
 \cH_0 =  \Pi_\phi^2 + \Pi_\sigma^2    +  \tfrac12 (\nabla \phi)^2 + \tfrac12 (\nabla \sigma)^2 + \tfrac12 M^2 \phi^2 + \tfrac12 m^2 \sigma^2  \,,
\ee
and
\bea
 \cH_{\rm int} &=&  \frac{1}{(1-\xi^2)} \Bigl[ \xi^2\Bigl( \Pi_\phi^2 + \Pi_\sigma^2\Bigr) - \xi \, \Pi_\phi \Pi_\sigma \Bigr] +   \xi \, (\nabla \phi) \cdot (\nabla \sigma) \nn\\
 &=& - \xi \, \Pi_\phi \Pi_\sigma  +   \xi \, (\nabla \phi) \cdot (\nabla \sigma) + \cO(\xi^2)  \\
 &=&  - \cL_{\rm int} + \cO(\xi^2)\,. \nn
\eea

To leading order in $\xi$ the purity is again given by \pref{AppPurityMasterEqFNested}, with the Wightman functions required in this example (for a system initially prepared in the vacuum state) given by
\bea
   W_{\mu\nu'}(x,x') &=& \langle 0 | \partial_\mu \sigma(x) \, \partial_\nu \sigma(x') | 0 \rangle 
   = \partial_\mu \partial_{\nu'} W^0(x,x') \\ \hbox{and} \quad
   \cW_{\mu\nu'}(x,x') &=&\xi^2  \langle 0 | \partial_\mu \phi(x) \, \partial_\nu \phi(x') | 0 \rangle = \xi^2 \partial_\mu \partial_{\nu'} \cW^0(x,x')  \,,
\eea
where $W^0(x,x')$ and $\cW^0(x,x')$ are the Wightman functions for $\sigma$ and $\phi$ respectively. In momentum space eq.~\pref{cWtocWk} becomes
\bea
   W^{\mu\nu'}_\bmk(t,t') &=& k^\mu k^\nu W^0_\bmk(t,t') \qquad \hbox{with} \qquad k^0 = \omega_\sigma(k) \nn\\
   \hbox{and} \quad
   \cW^{\mu\nu'}_\bmk(t,t') &=& k^\mu k^\nu \cW^0_\bmk(t,t') \qquad \hbox{with} \qquad k^0 = \omega_\phi(k) \,,
\eea
where $\omega_\sigma(k)$ and $\omega_\phi(k)$ are the appropriate dispersion relations and $W^0_{\bmk}$ and $\cW^0_\bmk$ are given by \pref{AppFlatWk}. 
 
Using this in eq.~\pref{PurityMasterEqFNested} then leads to 
\be  \label{AppPurityMasterEqFflatkin}
  \gamma_\bmk^s(t)   =  1  -  \frac{2\xi^2 (\omega_\sigma \omega_\phi + k^2)^2}{\omega_\sigma \omega_\phi(\omega_\sigma +\omega_\phi)^2} \, \sin^2 \Bigl[\tfrac12 (\omega_\sigma +\omega_\phi)(t-t_0) \Bigr]   \qquad \hbox{(sudden approximation)} \,,
\ee
where the approximate equality takes $M$ much larger than $m$ and $k$. The time average becomes
\be \label{AppPurityMasterEqFflatsecularkin}
  \gamma_\bmk^a(t) = 1 - \frac{\xi^2 (\omega_\sigma \omega_\phi + k^2)^2}{\omega_\sigma \omega_\phi(\omega_\sigma +\omega_\phi)^2}   \qquad \hbox{(adiabatic approximation)} \,.
\ee
This large-$M$ limit of these expressions of interest in the main text are given by
\be  \label{AppPurityMasterEqFflatkin2}
  \gamma_\bmk^s(t)   
  \simeq  1  -  \frac{2\xi^2\omega_\sigma}{ M} \, \sin^2 \Bigl[\tfrac12 M (t-t_0) \Bigr]  \qquad \hbox{(sudden approximation)} \,,
\ee
while the adiabatic result instead is
\be \label{AppPurityMasterEqFflatsecularkin2}
   \gamma_\bmk^a(t)   \simeq  1  -   \frac{\xi^2\omega_\sigma}{ M} \qquad \hbox{(adiabatic approximation)}\,.
\ee

\subsubsection{Nonderivative cubic couplings}

The next example uses the minimal interaction coupling the two fields in the scalar potential assuming invariance under $\phi \to - \phi$,
\be
\cL_{\rm int} = - \mfg   \, \phi^2 \sigma \,,
\ee
for coupling constant $\mfg$. Within this example the purity obtained by tracing out the $\phi$ field is again given by \pref{AppPurityMasterEqFNested} for the operators $\cO_{\rm sys} = \sigma$ and $\cO_{\rm env} = \mfg \, \phi^2$. 

Preparing the fields in their perturbative vacuum allows us to use \pref{AppFlatWk} to evaluate $W_\bmk$. The environmental correlator in this case 
is evaluated in \cite{Burgess:2024eng} and leads to the following rate of change for the purity
\be \label{AppPurityChangeMasterEqFSum2}
  \partial_t \gamma_\bmk = - \frac{2}{\omega_\sigma}  \int_{t_0}^t \exd s  \; \hbox{Re}\left[  e^{-i\omega_\phi (t-s)} \cW_{-\bmk}(t,s) \right] \,,
\ee
where the Wightman function for the operator $\mfg \phi^2$ is
\be  \label{AppFlatMassivecWk}
\cW_{\bf{k}}(t,s)
 =   \frac{i\mfg^2}{2(2\pi)^2k(t-s)} \int_0^\infty \frac{p \, \exd p}{\omega _\phi}\, e^{-i \omega _\phi (t-s)} \left[ e^{-i \omega _+(t-s)} - e^{-i \omega _-(t-s)} \right] \,,
\ee
with $\omega_\pm := \sqrt{(p \pm k)^2 + M^2}$. The small negative imaginary part that is required of $t-s$ for the Wightman function ensures the convergence of the $p$ integration. In the limit $k,m \ll M$ of most interest in the main text the purity evolution becomes
\be \label{AppPurityChangeMasterEqFSum2x}
  \partial_t \gamma_\bmk \simeq  \frac{\mfg^2}{4\pi^2\omega_\sigma} \int_1^\infty \frac{\exd u}{u^2} \sqrt{u^2-1} \; \hbox{Im}\left[  e^{-2i u M (t-t_0)} \right] \,.
\ee

The integrated purity is calculable in the sudden approximation in terms of Struve and Bessel functions \cite{Burgess:2024eng}, while turning on the coupling $\mfg$ adiabatically instead leads to the time-independent result
\be \label{AppPurityMassiveEnvEqFSum2xsecular}
  \gamma_\bmk^a \simeq
  1  -   \frac{\mfg^2}{16\pi^2\omega_\sigma M} \int_1^\infty \frac{\exd u}{u^3} \sqrt{u^2-1}  = 1  -   \frac{\mfg^2}{64\pi\omega_\sigma M}  \,.
\ee

\subsubsection{Derivative cubic couplings}
\label{Appsec:DerivativeCubic}

Our final example assumes the derivative cubic interaction
\begin{eqnarray}\label{Appdim5Int}
 \cL_{\rm int} = - \tfrac12  \kappa \,  \partial_\mu \phi \,  \partial^\mu \phi \, \sigma \,,
\end{eqnarray}
where $\kappa$ is a coupling with dimension length. 

Again the presence of time derivatives requires first computing the interaction Hamiltonian before the purity for the field $\sigma$ can be calculated. In this case the canonical momenta are
\be
  \Pi_\sigma = \frac{\delta S}{\delta \dot \sigma} =  \dot \sigma \quad \hbox{and} \quad
  \Pi_\phi = \frac{\delta S}{\delta \dot \phi} =  \dot \phi \Bigl(1 + \kappa \sigma \Bigr) \,,
\ee 
which inverts to give $\dot\sigma = \Pi_\sigma$ and $\dot \phi = \Pi_\phi/(1 + \kappa \sigma)$. The Hamiltonian $\cH =  \Pi_\phi \, \dot \phi + \Pi_\sigma \dot \sigma - \cL = \cH_0 + \cH_{\rm int}$ again has $\cH_0$ given by \pref{AppH0canon}, but with interaction Hamiltonian  
\be
  \cH_{\rm int} = - \tfrac{1}{2} \frac{\kappa \sigma \Pi_\phi^2}{1 + \kappa \sigma} + \tfrac12 \kappa  (\nabla \phi)^2 \sigma = \tfrac12   \kappa \sigma \Bigl[ -\dot \phi^2( 1 + \kappa \sigma) + (\nabla \phi)^2 \Bigr]    = - \cL_{\rm int} + \cO(\kappa^2)\,.
\ee
The Wightman functions required in this example therefore are (at leading order) for the operators $\cO_{\rm sys} = \sigma$ and $\cO_{\rm env} = \frac12 \kappa \, \partial_\mu \phi \, \partial^\mu \phi$. 

Assuming the system and environment are prepared in the vacuum the system correlator is therefore the usual single-field function $W_\bmk$ from \pref{AppFlatWk} and the environmental correlator is
\begin{eqnarray}
 \cW_{\bmk}(t_1,t_2) & = & \tfrac14\kappa^2 \int \exd^3\bmx \; \bigg[ \langle 0 | (\partial\phi)^2(t_1,\bmx) (\partial\phi)^2(t_2,\mathbf{0}) | 0 \rangle - \langle 0 | (\partial\phi)^2(t_1,\bmx)  | 0 \rangle \langle 0 | (\partial\phi)^2(t_2,\mathbf{0}) | 0 \rangle \bigg] \; e^{-i \bmk \cdot \bmx} \nn\\
& = & \tfrac12\kappa^2 \int \frac{\exd^3\bmp}{(2\pi)^3} \int \frac{\exd^3\bmq}{(2\pi)^3} \; \delta^{(3)}(\bmp + \bmq - \bmk) \; \bigg[ \dot u_p(t_1) \dot u_q(t_1) \dot u_p^{\ast}(t_2) \dot u_q^{\ast}(t_2)  \\
&& \qquad \qquad  - (\bmp \cdot \bmq)\; \Big\{ \dot u_p(t_1) \dot u_q(t_1) u_p^{\ast}(t_2) u_q^{\ast}(t_2) + u_p(t_1) u_q(t_1) \dot u_p^{\ast}(t_2) \dot u_q^{\ast}(t_2) \Big\} \notag \\
&& \qquad \qquad \qquad \qquad\qquad \qquad + (\bmp \cdot \bmq)^2 u_p(t_1) u_q(t_1) u_p^{\ast}(t_2) u_q^{\ast}(t_2) \bigg] \,, \notag
\end{eqnarray}
where $u_p = (2\omega_\phi)^{-1/2} \, e^{-i \omega_\phi(p) t}$ denotes the $\phi$-field mode functions.

Integrating over the angles defining the direction of $\bmp$ and $\bmq$ and changing variables in the integration over the magnitudes such that
\begin{eqnarray}
p = \tfrac12\Bigl( P+Q \Bigr) \qquad \mathrm{and} \qquad q = \tfrac12 \Bigl(P-Q\Bigr)
\end{eqnarray}
leaves one with:
\begin{equation}
\cW_{k}(t_1,t_2) = \frac{\kappa^2}{ (8 \pi )^5 k} \int_{k}^\infty \exd P \int_0^k \exd Q\; \sfrac{ (P^2 - Q^2) \left( P^2 + Q^2 - 2k^2 - 4 \omega_+\omega_-\right)^2 }{ \omega_+ \omega_- } \; e^{ - i \left[ \omega_+ + \omega_- \right] (t_1 - t_2 )} \,,
\end{equation}
where $\omega_\pm := \sqrt{\frac14(P\pm Q)^2 + M^2}$.

Inserting these expressions into the perturbative purity \pref{AppPurityMasterEqFNested} and performing the time integrals assuming an adiabatic onset of the coupling $\kappa$ finally gives
\begin{equation} \label{AppPP3}
\gamma_{\bmk}^a = 1 - \frac{ \kappa^2}{  (8\pi)^5 k \omega_\sigma } \int_k^\infty \exd P \int_0^k \exd Q\; \frac{ (P^2 - Q^2) \left( P^2 + Q^2 - 2k^2 - 4 \omega_+ \omega_- \right)^2 }{ \omega_+ \omega_- \big( \omega_\sigma + \omega_+ + \omega_- \big)^2 } 
\end{equation}
where $\omega_\sigma = \omega_\sigma(k)$. Notice that these integrals converge in the UV because the integrand of \pref{PP3} has the following large-$P$ form
\be
  \frac{ (P^2 - Q^2) \left( P^2 + Q^2 - 2k^2 - 4 \omega_+ \omega_- \right)^2 }{ \omega_+ \omega_- \big( \omega_\sigma + \omega_+ + \omega_- \big)^2 }= \frac{16(k^2 + 2M^2 - Q^2)^2}{P^2} + \cO(P^{-3}) \,.
\ee
In the large $\phi$-mass limit of most interest in the main text, for which $M \gg k,m$, the purity is well-approximated by\footnote{Recall that calculational control when using a dimension-5 coupling like \pref{Appdim5Int} also requires $\kappa M \ll 1$.}
\begin{equation} \label{AppPPF}
\gamma_{\bmk}^a \simeq 1 - \frac{ \kappa^2 M^3 }{ 1024 \pi^4 \omega_{\sigma} } \qquad \hbox{(adiabatic onset)} \ .
\end{equation}

\section{Schr\"odinger-picture wavefunctional calculations}
 \label{App:SchrodingerPicture}
 
For a complementary view of the purity calculations, we include here a discussion of how the purity, and more generally the reduced density matrix for the system, can be calculated within the Schrödinger field theory formalism (see \cite{Albrecht:2014aga} and references therein for more details). We focus here on the flat space case, since we can make use of the fact that an FRW spacetime is conformal to flat space (when the spatial sections are flat) and rescale the fields so that we can use the flat space treatment, albeit at the expense of making the couplings time dependent. 

Recall that in the Schrödinger picture we use a wavefunctional $\Psi$ that is a function of field configurations $\phi\left(\cdot\right)$  given at a time $t$. The time evolution of the system is then driven by the Schrödinger equation
\begin{equation}
i\partial_t \langle \phi\left(\cdot\right)|\Psi(t)\rangle = H\left[-i\frac{\delta}{\delta \phi\left(\cdot\right)}, \phi\left(\cdot\right) \right]\langle \phi\left(\cdot\right)| \Psi(t)\rangle
\end{equation}
If the problem at hand has spatial translations as a symmetry, we can describe the field configuration at a given time $t$ via (time independent) modes $\left\{\phi_{\bmk}\right\}$ and the relevant Schrödinger equation now takes the form:
\begin{equation}
i\partial_t \langle \left\{\phi_{\bmk}\right\}|\Psi(t)\rangle = H\left[\left\{-i\frac{\delta}{\delta \phi_{\bmk}}\right\},\left\{\phi_{\bmk}\right\} \right]\langle\left\{\phi_{\bmk}\right\}| \Psi(t)\rangle,
\end{equation}
where we've expanded our fields in a spatial box of volume $V$ as
\begin{equation}
\phi(\bmx) = \frac{1}{\sqrt{V}} \sum_{\bmk} \phi_{\bmk} e^{-i \bmk\cdot \bmx}.
\end{equation}
The generalization to more than one field is clear, and is what we will need in the sequel. 

Given a pure state wavefunctional $\Psi\left[\phi,\sigma,t\right]$ for two fields $\phi$ and $\sigma$, the reduced density matrix for the system described by $\sigma$ is given by tracing out the environmental field $\phi$:
\begin{equation}
\label{appeq:reduceddensitymatrix}
\langle \sigma \left| \varrho_{\mathrm{red}}(t)\right| \tilde{\sigma}\rangle = \int \mathcal{D}\phi\  \langle \phi, \sigma  |\Psi(t)\rangle\langle \Psi(t) | \phi,\tilde{\sigma}\rangle.
\end{equation}
From this relation, we can compute the purity as:
\begin{equation}
\label{eq:SpicPurity}
\gamma =\mathrm{Tr}\left(\varrho_{\mathrm{red}}^2\right)= \int \mathcal{D}\sigma\ \mathcal{D}\tilde{\sigma}\ \langle \sigma \left| \varrho_{\mathrm{red}}(t)\right| \tilde{\sigma}\rangle\langle  \tilde{\sigma} \left| \varrho_{\mathrm{red}}(t)\right| \sigma\rangle=\int \mathcal{D}\sigma\ \mathcal{D}\tilde{\sigma}\ \left|  \langle \sigma \left| \varrho_{\mathrm{red}}(t)\right| \tilde{\sigma}\rangle\right|^2.
\end{equation}

We now have the formalism in place to apply to the models considered in the main text.

\subsection{Decoherence from field mixing}
\label{appsubsec:mixing}

We'll focus on the case where the mixing comes from the mass matrix (as opposed to the kinetic terms) and the Hamiltonian can be written as:
\begin{equation}
\label{appeq:mixHam}
H_{\mathrm{mix}} = \int d^3 x\ \left\{\frac{\Pi_{\phi}^2}{2}+\frac{\Pi_{\sigma}^2}{2}+\frac{1}{2} \left(\nabla \phi\right)^2+\frac{1}{2} \left(\nabla \sigma\right)^2+\frac{1}{2} M^2 \phi^2+\frac{1}{2} m^2 \sigma^2+\mu^2 \phi\ \sigma\right\},
\end{equation}
where $\Pi_{\phi}$, $\Pi_{\sigma}$ are the canonically conjugate momenta to $\phi$, $\sigma$ respectively. We can rewrite this in terms of the modes and their canonical momenta, $H_{\mathrm{mix}} = \sum_{\bmk} H_{\bmk}$
with
\begin{equation}\label{appeq:mixHammodes}
    H_{\bmk} = \frac{\Pi^{\phi}_{\bmk} \Pi^{\phi}_{-\bmk}}{2}+\frac{\Pi^{\sigma}_{\bmk} \Pi^{\sigma}_{-\bmk}}{2}+\frac{1}{2} \omega_{k,\phi}^2 \phi_{\bmk}\phi_{-\bmk}+\frac{1}{2} \omega_{k,\sigma}^2 \sigma_{\bmk}\sigma_{-\bmk}+\frac{\mu^2}{2}\left( \phi_{\bmk} \sigma_{-\bmk}+\sigma_{\bmk}\phi_{-\bmk}\right)
\end{equation}
and
\begin{equation}
    \omega_{k,\phi}^2 = k^2+M^2,\qquad  \omega_{k,\sigma}^2 = k^2+m^2.
\end{equation}
In the Schrödinger picture, 
\begin{equation}
\label{appeq:conjmomenta}
\Pi^{\phi}_{\bmk} = -i\frac{\delta}{\delta \phi_{-\bmk}},
\end{equation}
and likewise for $\sigma$. 

Given the quadratic form of the Hamiltonian, it is natural to use a Gaussian anzatz for the wavefunctional
\begin{equation}
\label{appeq:mixwWFmodes}
\begin{split}
 \langle \left\{\phi_{\bmk}\right\}, \left\{\sigma_{\bmk}\right\}|\Psi(t)\rangle &= N(t) \exp\left(-\frac{\mathcal{K}\left[ \left\{\phi_{\bmk}\right\}, \left\{\sigma_{\bmk}\right\};t\right]}{2}\right)
\end{split}
\end{equation}
with 
\begin{equation}
    \mathcal{K}\left[\left\{\phi_{\bmk}\right\}, \left\{\sigma_{\bmk}\right\}; t\right] = \sum_{\bmq} \left\{\Gamma^{2,\phi}_q(t) \phi_{\bmq}\phi_{-\bmq}+\Gamma^{2,\sigma}_q(t) \sigma_{\bmq}\sigma_{-\bmq}+\Gamma^{2, \mathrm{mix}}_q(t)\left(\phi_{\bmq}\sigma_{-\bmq}+\phi_{-\bmq}\sigma_{\bmq} \right)\right\}.
\end{equation}
The pairing between $\bmq$ and $-\bmq$ in the modes enforces translational invariance, while rotational invariance dictates that the kernels $\Gamma^{2,\phi}_q(t),\Gamma^{2,\sigma}_q(t),\Gamma^{2,\mathrm{mix}}_q(t)$ depend only on the magnitude $q$ of $\bmq$. 
 
 One demand we have to impose on the state $|\Psi(t)\rangle$ is that it be normalizable. In the field representation, this will imply that
 \begin{equation}
 |N(t)|^2\int \mathcal{D}\sigma \mathcal{D}\phi\ |N(t)|^2 \exp\left(-\mathcal{K}_R\right)=1,
\end{equation}
where the subscript $R$ denotes the real part. In terms of the modes, if we make use of the fact that for real fields $\phi_{-\bmk}$ is the complex conjugate to $\phi_{\bmk}$ (and likewise for $\sigma$) the normalization condition becomes:
\begin{equation}
\label{appeq:modenorm}
\prod_{\bmk>0}\ \int |N_k(t)|^2\ \mathcal{D}^2 \sigma_{\bmk}\mathcal{D}^2 \phi_{\bmk}\ \exp\left(-2\left(\begin{array}{cc}\sigma_{\bmk} & \phi_{\bmk}\end{array}\right)\left(\begin{array}{cc}\Gamma^{2,\sigma}_{k,R}(t) & \Gamma^{2,\mathrm{mix}}_{k,R}(t) \\   \Gamma^{2,\mathrm{mix}}_{k,R}(t) & \Gamma^{2,\phi}_{k,R}(t)\end{array}\right)\left(\begin{array}{c}\sigma_{\bmk} \\ \phi_{\bmk}\end{array}\right)\right) =1.
\end{equation}
In writing eq.~\eqref{appeq:modenorm} we made use of the fact that the Gaussian wavefunctional factorizes in terms of the modes and that $\pm \bmk$ give equal contributions in the mode sum so as to factorize the normalization integral and restrict the product over wavenumbers to the ``positive'' half of them; this accounts for the factor of $2$ in the exponent. Also, the measure $\mathcal{D}^2 \phi_{\bmk}\equiv \mathcal{D}\mathrm{Re}\phi_{\bmk}\mathcal{D}\mathrm{Im}\phi_{\bmk}$ and likewise for $\sigma$. 

In order for the integrals over the modes to be finite, \emph{both} eigenvalues of the matrix in the exponent of eq.~\eqref{appeq:modenorm} must be positive. This can be fulfilled by requiring that the trace and the determinant both be positive, which implies:
\begin{subequations}
\label{appeq:positivityofeigens}
\begin{align}
& \Gamma^{2,\sigma}_{k,R}(t)>0,\qquad \Gamma^{2,\phi}_{k,R}(t)>0,\qquad  \Gamma^{2,\sigma}_{k,R}(t)\Gamma^{2,\phi}_{k,R}(t)-\Gamma^{2,\mathrm{mix}}_{k,R}(t)^2>0.
\end{align}
\end{subequations}

 Using the mode Hamiltonian in eq.~\eqref{appeq:mixHammodes} and the wavefunctional above in the Schrödinger equation, we can match powers of the modes on either side in order to generate differential equations for the kernels:
 \begin{subequations}
 \label{appeq:kernels}
 \begin{align}
 i\frac{\dot{N}(t)}{N(t)} &= \sum_{\bmk} \frac{\Gamma^{2,\phi}_k(t)+\Gamma^{2,\sigma}_k(t)}{2}\\
 i\dot{\Gamma}^{2,\phi}_k(t)&=\left(\Gamma^{2,\phi}_k(t)\right)^2+\left(\Gamma^{2,\mathrm{mix}}_k(t)\right)^2-\omega_{k,\phi}^2\\
 i\dot{\Gamma}^{2,\sigma}_k(t)&=\left(\Gamma^{2,\sigma}_k(t)\right)^2+\left(\Gamma^{2,\mathrm{mix}}_k(t)\right)^2-\omega_{k,\sigma}^2\\
 i\dot{\Gamma}^{2,\mathrm{mix}}_k(t)&=\Gamma^{2,\mathrm{mix}}_k(t)\left(\Gamma^{2,\sigma}_k(t)+\Gamma^{2,\phi}_k(t)\right)-\mu^2
 \end{align}
 \end{subequations}
We can recast eqs.~\eqref{appeq:kernels} into a more useful form by means of the so-called Ricatti trick: set 
\begin{equation}
\label{appeq:ricatti}
\Gamma^{2,\phi}_k(t)=-i \frac{\dot{u}_k(t)}{u_k(t)},\qquad \Gamma^{2,\sigma}_k(t)=-i \frac{\dot{v}_k(t)}{v_k(t)}. 
\end{equation}
Then we have
\begin{subequations}
 \label{appeq:kernelsricatti}
 \begin{align}
 i\frac{\dot{N}(t)}{N(t)} &=-\frac{i}{2} \sum_{\bmk}\left(\frac{\dot{u}_k(t)}{u_k(t)}+ \frac{\dot{v}_k(t)}{v_k(t)}\right)\\
\ddot{u}_k(t)+\omega_{k,\phi}^2 u_k(t) &=\left(\Gamma^{2,\mathrm{mix}}_k(t)\right)^2 u_k(t)\\
 \ddot{v}_k(t)+\omega_{k,\phi}^2 v_k(t) &=\left(\Gamma^{2,\mathrm{mix}}_k(t)\right)^2 v_k(t)\\
i\partial_t \left(u_k(t) v_k(t) \Gamma^{2,\mathrm{mix}}_k(t)\right) &= -\mu^2 u_k(t) v_k(t).
 \end{align}
 \end{subequations}
The last of eqs.~\eqref{appeq:kernelsricatti} is telling; if we interpret $\Gamma^{2,\mathrm{mix}}_k(t)$ as a measure of the entaglement between the system described by $\sigma$ and the environment $\phi$, then that equation tells us that if $\mu=0$ and we assume that the field are not entangled at the initial time $t_0$ (i.e. $\Gamma^{2,\mathrm{mix}}_k(t_0)=0$), the states will remain unentangled. Thus the mixing in the mass matrix acts as a source of entanglement which will be seen to be a prerequisite for generating a mixed reduced density matrix for the system $\sigma$.

Our description of the kernels in terms of modes $u_k,\ v_k$ together with the first of eqs.~\eqref{appeq:positivityofeigens} allows us to derive an interesting constraint on the modes. First note that $2 \Gamma^{2,\phi}_{k,R}(t)=i W[u_k(t), u^*_k(t)]\slash |u_k(t)|^2,\ 2 \Gamma^{2,\sigma}_{k,R}(t)=i W[v_k(t), v^*_k(t)]\slash |v_k(t)|^2$, where $W$ denotes the Wronskian of the modes indicated. Let's focus on the Wronskian for $u_k, u^*_k$. Because of the source on the right hand side of the mode equations in eq.~\eqref{appeq:kernelsricatti}, this Wronksian is not constant:
\begin{equation*}
\partial_t W[u_k(t),u^*_k(t)]= -2 i \mathrm{Im}\left(\left(\Gamma^{2,\mathrm{mix}}_k(t)\right)^2\right) |u_k(t)|^2. 
\end{equation*}
For initially unentangled states, this derivative vanishes at $t_0$ but not at later times. However, if, as we plan to do, we treat $\mu^2$ as a perturbation and thus $\left(\Gamma^{2,\mathrm{mix}}_k(t)\right)^2$ can be neglected compared to the harmonic term $\omega_{k,\phi}^2$ in the mode equations, then to lowest order n $\mu^2$ the Wronksian can be taken as \emph{constant}  and the positivity constraints on the kernels then imply that we should choose $i W[u_k(t), u^*_k(t)]>0$; in fact, we will normalize this to unity. Then at this order in perturbation theory, the modes satisfying the Wronksian condition are just:
\begin{equation}
\label{appeq:zerothordermodes}
u_k(t) = \frac{e^{i \omega_{k,\phi} t}}{\sqrt{2 \omega_{k,\phi}}}, \quad v_k(t) = \frac{e^{i \omega_{k,\sigma} t}}{\sqrt{2 \omega_{k,\sigma}}}, 
\end{equation}
and the last positivity constraint becomes 
\begin{equation}
 \omega_{k,\phi} \omega_{k,\sigma}- \Gamma^{2,\mathrm{mix}}_{k,R}(t)^2>0.
\end{equation}
Given the modes in eq.~\eqref{appeq:zerothordermodes}, we can solve the last equation in eq.~\eqref{appeq:kernelsricatti} for $\Gamma^{2,\mathrm{mix}}_{k,R}(t)$:
\begin{equation}
\label{appeq:zerothordermixkernel}
\Gamma^{2,\mathrm{mix}}_k(t) = \frac{i \mu^2}{ \omega_{k,\phi}+ \omega_{k,\sigma}}\left(1-\exp\left(-i\left(\omega_{k,\phi}+ \omega_{k,\sigma} \right)\left(t-t_0\right)\right)\right),
\end{equation}
where we understand that the time integration contour is to be shifted by $i\epsilon$ as discussed in the text.

We still need to compute the reduced density matrix for the system $\sigma$ and the resulting purity. For this Gaussian system, the different modes decouple from one another so we can compute the mode reduced density matrix as well as the mode purity. The reduced density matrix element $\langle \sigma_{\bmk}|\varrho_{\mathrm{red},k}(t)|\tilde{\sigma}_{\bmk}\rangle$ is given by tracing out $\phi_{\bmk}$:
\begin{equation}
\begin{split}
\langle \sigma_{\bmk}|\varrho_{\mathrm{red},k}(t)|\tilde{\sigma}_{\bmk}\rangle &= \int \mathcal{D}^2\phi_{\bmk}\ \langle \phi_{\bmk},\sigma_{\bmk}|\rho_{k}(t)|\tilde{\phi}_{\bmk},\tilde{\sigma}_{\bmk}\rangle = \exp\left(-\Gamma^{2,\sigma}_k(t) \sigma_{\bmk} \sigma_{-\bmk}-\Gamma^{2,\sigma *}_k(t) \tilde{\sigma}_{\bmk} \tilde{\sigma}_{-\bmk} \right) \mathcal{I}
\end{split}
\end{equation}
with 
\begin{equation}
\begin{split}
\mathcal{I} &= \int \mathcal{D}^2\phi_{\bmk}\ \exp\left(-2\ \Gamma^{2,\phi}_{k, R}\ \phi_{\bmk}\phi_{-\bmk}-\phi_{\bmk} J_{-\bmk}-\phi_{-\bmk} J_{\bmk}\right), \qquad
J_{\bmk} = \Gamma^{2,\mathrm{mix}}_k \sigma_{\bmk}+ \Gamma^{2,\mathrm{mix} *}_k \tilde{\sigma}_{\bmk}.
\end{split}
\end{equation}
We can compute $\mathcal{I}$ by completing the square, or equivalently by making noting that the saddle point approximation is exact here. Either way, the result is (up to a prefactor that we will subsume into the overall normalization of the reduced density matrix):
\begin{equation}
\mathcal{I} \propto \exp\left( \frac{J_{\bmk} J_{-\bmk}}{2 \Gamma^{2,\phi}_{k,R}}\right),
\end{equation}
which leads to
\begin{equation}
\label{appeq:redmatrix}
\begin{split}
\langle \sigma_{\bmk}|\varrho_{\mathrm{red},k}(t)|\tilde{\sigma}_{\bmk}\rangle & \propto \exp\left(-\alpha_k\ \sigma_{\bmk}\sigma_{-\bmk}-\alpha_k^*\ \tilde{\sigma}_{\bmk}\tilde{\sigma}_{-\bmk}+\beta_k\ \left(  \sigma_{\bmk}\tilde{\sigma}_{-\bmk}+\tilde{\sigma}_{\bmk} \sigma_{-\bmk}\right)\right)
\end{split}
\end{equation}
with 
\begin{equation}
\begin{split}
\alpha_k &= \Gamma^{2,\sigma}_k -\frac{\left(\Gamma^{2,\mathrm{mix}}_k\right)^2}{2 \Gamma^{2,\phi}_{k,R}}, \qquad
\beta_k = \frac{\left| \Gamma^{2,\mathrm{mix}}_k\right|^2}{2 \Gamma^{2,\phi}_{k,R}}.
\end{split}
\end{equation}
As expected, tracing out $\phi$ has left us with a reduced density matrix describing a mixed state with the amount of mixing controlled by the entanglement parameter $\Gamma^{2,\mathrm{mix}}_k$. We normalize this density matrix by setting $\tilde{\sigma}_{\bmk} = \sigma_{\bmk}$ and integrating over $ \sigma_{\bmk}$:
\begin{equation}
\label{appeq:redmatrixnorm}
1= \mathcal{N}_k \int \mathcal{D}^2\sigma_{\bmk} \exp\left(-2\left(\alpha_{k,R}-\beta_k\right)\ \sigma_{\bmk}\sigma_{-\bmk}\right)=\frac{\pi}{2} \left(\frac{ \mathcal{N}_k}{\alpha_{k,R}-\beta_k}\right),
\end{equation}
where $\mathcal{N}_k$ is the overall normalization of the reduced density matrix. Note that 
\begin{equation}
\alpha_{k,R}-\beta_k= \Gamma^{2,\sigma}_{k, R}-\frac{\left(\Gamma^{2,\mathrm{mix}}_{k,R}\right)^2}{\Gamma^{2,\phi}_{k,R}},
\end{equation}
which by the last positivity condition in eq.~\eqref{appeq:positivityofeigens} is positive; this ensures that the normalization integral converges. 
Thus we can finally write:
\begin{equation}
\label{appeq:finalredmatrix}
\langle \sigma_{\bmk}|\varrho_{\mathrm{red},k}(t)|\tilde{\sigma}_{\bmk}\rangle=\frac{2}{\pi} \left( \Gamma^{2,\sigma}_{k, R}-\frac{\left(\Gamma^{2,\mathrm{mix}}_k\right)^2}{ \Gamma^{2,\phi}_{k,R}}\right)\exp\left(-\alpha_k\ \sigma_{\bmk}\sigma_{-\bmk}-\alpha_k^*\ \tilde{\sigma}_{\bmk}\tilde{\sigma}_{-\bmk}+\beta_k\ \left(  \sigma_{\bmk}\tilde{\sigma}_{-\bmk}+\tilde{\sigma}_{\bmk} \sigma_{-\bmk}\right)\right).
\end{equation}

We can use eq.~\eqref{appeq:finalredmatrix} to obtain the purity of the state using the result of eq.~\eqref{eq:SpicPurity}. First we note that since the reduced density matrix factorizes in this example so will the purity:
\begin{equation}
\gamma = \prod_{\bmk}\ \gamma_{\bmk},
\end{equation}
so that it is sufficient to just focus on a particular $\bmk$ mode:
\begin{equation}
\label{appeq:mixpurity1}
\begin{split}
\gamma_{\bmk} &= \int \mathcal{D}^2\sigma_{\bmk}\mathcal{D}^2\tilde{\sigma}_{\bmk}\ \left|\langle \sigma_{\bmk}|\varrho_{\mathrm{red},k}(t)|\tilde{\sigma}_{\bmk}\rangle\right|^2\\
=& \left(\frac{2}{\pi}\right)^2 \left(\alpha_{k,R}-\beta_k\right)^2  \int \mathcal{D}^2\sigma_{\bmk}\mathcal{D}^2\tilde{\sigma}_{\bmk}\ \exp\left(-2\alpha_{k,R}\ \left(\sigma_{\bmk}\sigma_{-\bmk}+ \tilde{\sigma}_{\bmk}\tilde{\sigma}_{-\bmk}\right)+2\beta_k\ \left( \sigma_{\bmk}\tilde{\sigma}_{-\bmk}+\tilde{\sigma}_{\bmk} \sigma_{-\bmk}\right)\right).
\end{split}
\end{equation}
The integrand is a Gaussian in $\sigma_{\bmk},\tilde{\sigma}_{\bmk}$, whose quadratic form matrix is hermitian with eigenvalues $\alpha_{k, R}\pm\beta_k$. Changing variables via a unitary transformation allows us to perform the integral to find
\begin{equation}
\label{appeq:mixpurity2}
\gamma_{\bmk} = \frac{\alpha_{k,R}-\beta_k}{\alpha_{k,R}+\beta_k}=1-\frac{2 \beta_k}{\alpha_{k,R}+\beta_k}\leq 1.
\end{equation}
These results are exact for a quadratic system. However, we can use the fact that we are treating $\mu^2$ as a perturbatively small quantity and hence $\Gamma^{2,\mathrm{mix}}_k$ and thus $\beta_k$ as parametrically smaller than $\alpha_{k,R}$ to write an approximate form of the purity:
\begin{equation}
\label{appeq:mixpurityapprox}
\gamma_{\bmk} \simeq 1-2 \frac{\beta_k}{\alpha_{k,R}} \simeq 1- \frac{\left| \Gamma^{2,\mathrm{mix}}_k\right|^2}{ \Gamma^{2,\phi}_{k,R} \Gamma^{2,\sigma}_{k,R}}=1- \frac{\left| \Gamma^{2,\mathrm{mix}}_k\right|^2}{ \omega_{k,\phi} \omega_{k,\sigma}}.
\end{equation}
We can finish the computation of $\gamma_{\bmk}$ by using eq.~\eqref{appeq:zerothordermixkernel}. We have:
 \begin{equation}
 \begin{split}
\left| \Gamma^{2,\mathrm{mix}}_k\right|^2&=\frac{\mu^4}{ \left(\omega_{k,\phi}+ \omega_{k,\sigma}\right)^2}\left|1-\exp\left(-i\left(\omega_{k,\phi}+ \omega_{k,\sigma} \right)\left(t-t_0\right)\right)\right|^2\\
& =\frac{4 \mu^4}{ \left(\omega_{k,\phi}+ \omega_{k,\sigma}\right)^2} \sin^2\left(\frac{1}{2}\left(\omega_{k,\phi}+ \omega_{k,\sigma} \right)\left(t-t_0\right)\right)
\end{split}
\end{equation}
If we time average this result, we arrive at
\begin{equation}
\label{appeq:mixkerneltimeaveraged}
\left| \Gamma^{2,\mathrm{mix}}_k\right|^2 = \frac{2 \mu^4}{ \left(\omega_{k,\phi}+ \omega_{k,\sigma}\right)^2}. 
\end{equation}
This result is twice what was obtained using the correlation function method. The reason is that this calculation captures contributions both from $\bmk$ as well as $-\bmk$. These are equal so we have to take the square root of our result and to lowest order in $\mu^2$, that requires taking only half of the result in eq.~\eqref{appeq:mixkerneltimeaveraged}. 

\subsection{Decoherence from cubic interactions}
\label{subsec:cubicint}

Another case treated in the main text is that of a cubic interaction between the system field $\sigma$ and the environment $\phi$. The possible such cubic interactions consistent with $\phi\leftrightarrow -\phi$ as well as time reversal symmetries are:
\begin{equation}
\label{appeq:cubicinteractionsLag}
-\mathcal{L}_{\mathrm{int}}= \frac{\mathfrak{g}}{2} \phi^2 \sigma-\frac{1}{2 } \left(\kappa_t \dot{\phi}^2 - \kappa_s \left(\nabla\phi\right)^2\right)\sigma.
\end{equation}
Here $\mathfrak{g}$ has mass dimension $1$ while $\kappa_{t,s}$ have dimension $-1$. To lowest order in perturbation theory, this corresponds to an interaction Hamiltonian density:
\begin{equation}
\label{appeq:cubicinteractionsHam}
\mathcal{H}_{\mathrm{int}}= \frac{\mathfrak{g}}{2} \phi^2 \sigma-\frac{1}{2 }\left(\kappa _t \left(\Pi^{\phi}\right)^2 - \kappa_s \left(\nabla\phi\right)^2\right)\sigma.
\end{equation}

\subsubsection{Perturbative framework}
\label{subsubsec:pertframe}
Unlike the mass mixing case treated above, here we cannot solve the Schrödinger equation exactly; a perturbative expansion must be constructed. We do this in some generality in this subsection and then apply it to the cubic interactions above.

Thus suppose our Hamiltonian takes the form $H= H_0+g H_{\mathrm{int}}$, where $g$ is a generic (small) coupling we can perturb in. Furthermore, suppose that we construct a wavefunctional $\Psi_0[\phi,\sigma;t]$ such that:
\begin{equation}
\label{appeq:zerothorderwf}
i\partial_t \Psi_0[\phi,\sigma;t] = H_0 \Psi_0[\phi,\sigma;t]. 
\end{equation}
Then to leading order in $g$, we write the full wavefunctional $\Psi[\phi,\sigma;t]$ as:
\begin{equation}
\label{appeq:pertubwf}
\Psi[\phi,\sigma;t] = \left(1-g\Delta[\phi,\sigma;t]\right) \Psi_0[\phi,\sigma;t].
\end{equation}
Inserting this into the Schrödinger equation and matching terms in $g$ then yields:
\begin{equation}
\label{appeq:firstordercorrection}
\left(i\partial_t\Delta\right)\ \Psi_0[\phi,\sigma;t] = \left[H_0, \Delta\right]\ \Psi_0[\phi,\sigma;t]- H_{\mathrm{int}} \Psi_0[\phi,\sigma;t].
\end{equation}
As we will see below, once an ansatz is chosen for the perturbation $\Delta[\phi,\sigma;t]$, this equation can be solved for the parameters appearing in the ansatz.

Given this form of the state, we can trace out $\phi$ to compute the reduced density matrix within this perturbative scheme:
\begin{equation}
\label{appeq:redmatpert}
\langle\sigma \left| \varrho_{\mathrm{red}}(t)\right| \tilde{\sigma}\rangle = \int \mathcal{D}\phi\  \left(1-g\Delta[\phi,\sigma;t]\right) \Psi_0[\phi,\sigma;t] \Psi^*_0[\phi,\tilde{\sigma};t]  \left(1-g\Delta^*[\phi,\tilde{\sigma};t]\right).
\end{equation}
We also have to enforce that the full state be normalizable. This requires
\begin{equation}
\label{appeq:pertstatenorm}
1 = \int \mathcal{D}\phi\ \mathcal{D}\sigma\  \left(1-g\Delta[\phi,\sigma;t]\right)  \left(1-g\Delta^*[\phi,\sigma;t]\right) \Psi_0[\phi,\sigma;t] \Psi^*_0[\phi,\sigma;t].
\end{equation}
Enforcing this order by order in $g$ tells us that
\begin{subequations}
\begin{align}
1 &= \int \mathcal{D}\phi\ \mathcal{D}\sigma\  \Psi_0[\phi,\sigma;t] \Psi^*_0[\phi,\sigma;t]\\
0 & =\int \mathcal{D}\phi\ \mathcal{D}\sigma\ \mathrm{Re}\left(\Delta[\phi,\sigma;t] \right)\Psi_0[\phi,\sigma;t] \Psi^*_0[\phi,\sigma;t]
\end{align}
\end{subequations}
There's an important point to note about eq.~\eqref{appeq:redmatpert}. We have only expanded the state to order $g$, but, unlike the normalization condition, we will expand the reduced density matrix to $\mathcal{O}(g^2)$. The reason for this is that this will be the leading contribution to the \emph{mixing} term in the density matrix. There will be other $\mathcal{O}(g^2)$ when we expand the state further, but in essence, $\mathcal{O}(g^2)$ in eq.~\eqref{appeq:redmatpert} is a correction to $0$, so should be kept. 

We can use the general perturbative framework discussed above to the interaction Hamiltonian in eq.~\eqref{appeq:cubicinteractionsHam}. In order to construct the relevant ansatzë for $\Psi_0[\phi,\sigma;t] , \Delta[\phi,\sigma;t]$, we first note that they should preserve the $\phi\leftrightarrow -\phi$ invariance of the interaction terms. Second, we note that these interactions contribute to the $\sigma$ tadpole. Thus we both need to add a tadpole term to the interaction Hamiltonian, 
\begin{equation}
\label{appeq:tadpole}
\mathcal{H}_{\mathrm{tadpole}} = -\lambda(t) \sigma,
\end{equation}
as well as a term in the state so that the expectation value of $\sigma$ vanish in the state:
\begin{equation}
\label{appeq:tadpolestate}
0=\langle \sigma \rangle = \mathrm{Tr}\left( \varrho_{\mathrm{red}}(t) \sigma\right)=\int \mathcal{D}\phi\ \mathcal{D}\sigma\ \sigma  \left(1-g\Delta[\phi,\sigma;t]\right)\ \left(1-g\Delta^*[\phi,\sigma;t]\right) \Psi_0[\phi,\sigma;t] \Psi^*_0[\phi,\sigma;t].
\end{equation}

Given these requirements on the state, we write
\begin{subequations}
\begin{align}
\Psi_0[\left\{\phi_{\bmk}\right\},\left\{\sigma_{\bmk}\right\};t]&= N_0(t)\exp\left(-\frac{1}{2} \sum_{\bmq}\left\{ \Gamma^{2,\phi}_q(t) \phi_{\bmq}\phi_{-\bmq}+ \Gamma^{2,\sigma}_q(t) \sigma_{\bmq}\sigma_{-\bmq}\right\}\right)\\
\Delta[\phi,\sigma;t]&= \sum_{\bmq} \Gamma^1_{\bmq}(t)\ \sigma_{-\bmq}+\frac{1}{2}\sum_{\bmq_1,\bmq_2,\bmq_3} \Gamma^3_{\bmq_1,\bmq_2,\bmq_3}(t)\ \phi_{\bmq_1}\ \phi_{\bmq_2}\ \sigma_{\bmq_3},
\end{align}
\end{subequations}
where translational invariance forces $\Gamma^1_{\bmq}(t)\propto \delta_{\bmq,\mathbf{0}},\ \Gamma^3_{\bmq_1,\bmq_2,\bmq_3}(t)\propto \delta_{ \bmq_1+ \bmq_2+ \bmq_3, \mathbf{0}}$. In order to calculate the reduced density matrix, we first compute the equations of motion for these kernels, then discuss the implications of tadpole cancellation for the kernels and finally trace out the environmental field to arrive at the reduced density matrix.

\subsubsection{Schrödinger equation for the kernels}
\label{subsubsec:seqn}

Having chosen forms for the zeroth order wavefunctional as well as its perturbation $\Delta[\phi,\sigma;t]$, we can use the free Hamiltonian given by $H_0 = \sum_{\bmk} H_{\bmk}$ with
\begin{equation}
\label{appeq:freeHam}
\begin{split}
H_{\bmk} &= \frac{\Pi^{\phi}_{\bmk} \Pi^{\phi}_{-\bmk}}{2}+\frac{\Pi^{\sigma}_{\bmk} \Pi^{\sigma}_{-\bmk}}{2}+\frac{1}{2} \omega_{k,\phi}^2 \phi_{\bmk}\phi_{-\bmk}+\frac{1}{2} \omega_{k,\sigma}^2 \sigma_{\bmk}\sigma_{-\bmk}
\end{split}
\end{equation}
and
\begin{equation}
    \omega_{k,\phi}^2 = k^2+M^2,\qquad  \omega_{k,\sigma}^2 = k^2+m^2
\end{equation}
in Eq.~\eqref{appeq:firstordercorrection}. We then match the various powers of the field modes on either side of eq.~\eqref{appeq:firstordercorrection} to arrive at the equations for the time evolution of the kernels. Doing this yields
\begin{equation}
\label{appeq:cubickernels}
\begin{split}
i\dot{\Gamma}_k^{2,\phi} &=\left(\Gamma_k^{2,\phi}\right)^2-\omega_{k ,\phi}^2\\
i\dot{\Gamma}_k^{2,\sigma} &=\left(\Gamma_k^{2,\sigma}\right)^2-\omega_{k ,\sigma}^2\\
i\dot{\Gamma}^1&= \Gamma^1 \Gamma^{2,\sigma}_{k=0} +\lambda \sqrt{V}+ \sum_{\bmq} \left[\frac{\kappa_t}{2\sqrt{V}} \Gamma^{2,\phi}_q -\frac{1}{2} \Gamma^3_{\bmq,-\bmq,\mathbf{0}}\right]\\
i\dot{\Gamma}^3_{\bmq_1,\bmq_2\bmq_3}&=\Gamma^3_{\bmq_1,\bmq_2\bmq_3}\left( \Gamma^{2,\phi}_{q_1}+\Gamma^{2,\phi}_{q_2}+\Gamma^{2,\sigma}_{q_3}\right)+\frac{1}{\sqrt{V}}\left(-2 \mathfrak{g}+\kappa_s \left( \bmq_1\cdot \bmq_2\right)-\kappa_t \Gamma_{q_1}^{2,\phi}\Gamma_{q_2}^{2,\phi}\right)\delta_{\bmq_1+\bmq_2+\bmq_3,\mathbf{0}}
\end{split}
\end{equation}
where we have written $\Gamma^1_{\bmk}$ as $\delta_{\bmk,\mathbf{0}} \Gamma^1(t)$.

As in the mixing case treated in sec.\ref{appsubsec:mixing} we can make use of the Ricatti trick of eq.~\eqref{appeq:ricatti} and rewrite eqs.~\eqref{appeq:cubickernels} as
\begin{subequations}
\label{appeq:kernelmodes}
\begin{align}
\ddot{u}_k+\omega_{k ,\phi}^2 u_k &=0\\
\ddot{v}_k+\omega_{k ,\sigma}^2 v_k &=0\\
i\partial_t\left(v_{k=0} \Gamma^1\right) &= v_{k=0}(t)\left\{ \lambda \sqrt{V}+ \sum_{\bmq} \left[\frac{\kappa_t}{2\sqrt{V}} \Gamma^{2,\phi}_q -\frac{1}{2} \Gamma^3_{\bmq,-\bmq,\mathbf{0}}\right]\right\}\\
i\partial_t\left(u_{q_1} u_{q_2} v_{q_3} \Gamma^3_{\bmq_1,\bmq_2\bmq_3}\right) &=\frac{1}{\sqrt{V}}\left[\left(-2 \mathfrak{g}+\kappa_s \left( \bmq_1\cdot \bmq_2\right)\right)u_{q_1} u_{q_2} v_{q_3} -\kappa_t \dot{u}_{q_1}\dot{u}_{q_2} v_{q_3} \right]\delta_{\bmq_1+\bmq_2+\bmq_3,\mathbf{0}}\nonumber\\
&= \frac{1}{\sqrt{V}}\left[-2 \mathfrak{g}+\kappa_s \left( \bmq_1\cdot \bmq_2\right) +\kappa_t\  \omega_{q_1,\phi}\ \omega_{q_2,\phi} \right]\left(u_{q_1} u_{q_2} v_{q_3}\right)\delta_{\bmq_1+\bmq_2+\bmq_3,\mathbf{0}},
\end{align}
\end{subequations}
where we made use of $\dot{u}_k = i \omega_{k,\phi} u_k,\ \dot{v}_k = i \omega_{k,\sigma} v_k$ to obtain the last equality.

\subsubsection{Tadpole cancellation}
\label{subsubsec:tadpole}

Let's turn now to the cancellation of the tadpole term. Taking the coupling $g$ to refer to any of $\mathfrak{g},\kappa_t,\ \kappa_s$, we compute:
\begin{equation}
\begin{split}
\langle \sigma_{\bmq} \rangle &= \int \left[\prod_{\bmk} \mathcal{D}^2\phi_{\bmk}\ \mathcal{D}^2\sigma_{\bmk}\right]\  \sigma_{\bmq} \left(1-g \left( \Delta[\left\{ \phi_{\bmk}\right\},\left\{\sigma_{\bmk}\right\};t]+ \Delta^*[\left\{ \phi_{\bmk}\right\},\left\{\sigma_{\bmk}\right\};t]\right)\right .\\
&+g^2 \left .   \left|\Delta[\left\{ \phi_{\bmk}\right\},\left\{\sigma_{\bmk}\right\};t]\right|^2\right) \left|\Psi_0[\left\{\phi_{\bmk}\right\},\left\{\sigma_{\bmk}\right\};t]\right|^2\\
\end{split}
\end{equation}
The only contribution comes from the $\mathcal{O}(g)$ terms. Using the Gaussian nature of $\Psi_0[\left\{\phi_{\bmk}\right\},\left\{\sigma_{\bmk}\right\};t]$, we can write the two-point functions
\begin{equation}
\label{appeq:twopoints}
\begin{split}
\langle \sigma_{\bmk} \sigma_{\bmq} \rangle_0 &\equiv  \int \left[\prod_{\vec{l}} \mathcal{D}^2\phi_{\vec{l}}\ \mathcal{D}^2\sigma_{\vec{l}}\right]\  \sigma_{\bmk} \sigma_{\bmq}  \left|\Psi_0[\left\{\phi_{\bmk}\right\},\left\{\sigma_{\bmk}\right\};t]\right|^2=\frac{\delta_{\bmk+\bmq,\mathbf{0}}}{\Gamma^{2,\sigma}_{k,R}}\\
\langle \phi_{\bmk} \phi_{\bmq} \rangle_0 &\equiv  \int \left[\prod_{\vec{l}} \mathcal{D}^2\phi_{\vec{l}}\ \mathcal{D}^2\sigma_{\vec{l}}\right]\  \phi_{\bmk} \phi_{\bmq} \left|\Psi_0[\left\{\phi_{\bmk}\right\},\left\{\sigma_{\bmk}\right\};t]\right|^2=\frac{\delta_{\bmk+\bmq,\mathbf{0}}}{\Gamma^{2,\phi}_{k,R}},
\end{split}
\end{equation}
which then allows us to write the $\mathcal{O}(g)$ tadpole contribution as:
\begin{equation}
\langle \sigma_{\bmq} \rangle = -\frac{\delta_{\bmq,\mathbf{0}}}{\Gamma^{2,\sigma}_{q,R}}\left[\Gamma^1_{\bmq=\mathbf{0}} +\frac{1}{2}\sum_{\bmk} \frac{\Gamma^3_{\bmk, -\bmk,\mathbf{0}}}{\Gamma^{2,\phi}_{k,R}} \right]+\mathrm{c.c}\equiv  -\frac{\delta_{\bmq,\mathbf{0}}}{\Gamma^{2,\sigma}_{q,R}}\left(T+T^*\right), 
\end{equation} 
so canceling the tadpole can be done by setting $T=0$. A question that comes up at this point is what the role of the tadpole term in the interaction Hamiltonian is. To answer this question, we can examine what condition arises from the requirement that if $T$ is set to zero at a given time, that it remains zero for \emph{all} times, i.e. that $\partial_t T=0$. This involves $\partial_t \Gamma^1$, which in turn brings in $\lambda$ together with other terms. The function $\lambda$ is then used to set $\partial_t T=0$.

\subsubsection{The reduced density matrix}
\label{subsubsec:reduceddensitymatrix}

We now have all the ingredients laid out to be able to compute the reduced density matrix for $\sigma$. We trace $\phi$ out of the pure state density matrix:
\begin{equation}
\label{eq:puredensity}
        \langle \phi,\sigma\left | \right . \rho(t)\left | \right . \Tilde{\phi}, \Tilde{\sigma} \rangle =\langle \phi, \sigma\left | \right. \Psi(t)\rangle \langle \Psi(t)\left | \right. \Tilde{\phi},\Tilde{\sigma}\rangle,
\end{equation}
by setting $\Tilde{\phi}=\phi$ and doing the path integral over $\phi$
This corresponds to calculating:
\begin{equation}
\label{eq:reducedtrace}
    \begin{split}
        \langle \sigma\left| \right . \varrho_{\mathrm{red}}(t) \left| \right . \Tilde{\sigma}\rangle &= \int \prod_{\bmk}\mathcal{D}^2\phi_{\bmk}\ \langle \phi, \sigma\left | \right. \Psi(t)\rangle \langle \Psi(t)\left | \right. \phi,\Tilde{\sigma}\rangle\\
        &= \int \prod_{\bmk}\mathcal{D}^2\phi_{\bmk}\ \left(1-\Delta\left[\phi,\sigma\right]\right)\left(1-\Delta^*\left[\phi,\Tilde{\sigma}\right]\right) \left |\Psi_G\left[\phi\right]\right |^2 \Psi_G\left[\sigma\right]\Psi_G^*\left[\Tilde{\sigma}\right]. 
    \end{split}
\end{equation}
In eq.~\eqref{eq:reducedtrace} we have factored the pure Gaussian part of the wavefunctional into separate $\phi$ and $\sigma$ dependent pieces
\begin{equation}
 \Psi_G\left[\phi\right]=N_G(t)\exp{\left(-\frac{1}{2} \sum_{\bmq}\Gamma^{2,\phi}_q \phi_{\bmq}\ \phi_{-\bmq}\right)},   
\end{equation}
and likewise for $\sigma$. We can open up the product 
\begin{equation*}
    \left(1-g \Delta\left[\phi,\sigma\right]\right)\left(1-g \Delta^*\left[\phi,\Tilde{\sigma}\right]\right)
\end{equation*} 
in powers of the generic perturbative coupling $g$. The leading term just gives $\Psi_G\left[\sigma\right]\Psi_G^*\left[\Tilde{\sigma}\right]$. The order $g$ term would give pieces proportional to a single mode $\sigma_{\bmk}$, which would induce a tadpole. We expect that the vanishing tadpole condition found above would eliminate this term and it does:
\begin{equation}
    \label{eq:ordergtadpole}
 \int \prod_{\bmk}\mathcal{D}^2\phi_{\bmk}\ \Delta\left[\phi,\sigma\right]\ \left |\Psi_G\left[\phi\right]\right |^2=\sigma_{\bmk=\mathbf{0}}\ \left(\Gamma^1+\frac{1}{2}\sum_{\bmq} \frac{\Gamma^3_{\bmq,-{\bmq},\mathbf{0}}}{\Gamma^{2,\phi}_{R,q}}\right)= \sigma_{\bmk=\mathbf{0}}\ T. 
\end{equation}
This means that the leading contribution from tracing out the environment will come at order $g^2$. The tadpole condition comes into play for this term as well. When we expand out $\Delta\left[\phi,\sigma\right]\ \Delta^*\left[\phi,\Tilde{\sigma}\right]$, we'll get the following terms:
\begin{subequations}
        \begin{align}
            t_1 &=\left(\sigma_{\bmq=\mathbf{0}}\ \Tilde{\sigma}_{\bmq=\mathbf{0}}\right) \Gamma^1 \Gamma^{1 *}\\
            t_2 &=\left(\sigma_{\bmk=\mathbf{0}}\ \Tilde{\sigma}_{\bmk=\mathbf{0}}\right) \frac{1}{2}\sum_{\bmq} \left\{\Gamma^1\ \frac{\Gamma^{3 *}_{\bmq,-{\bmq},\mathbf{0}}}{\Gamma^{2,\phi}_{R,q}}+\Gamma^{1 *}\ \frac{\Gamma^3_{\bmq,-{\bmq},\mathbf{0}}}{\Gamma^{2,\phi}_{R,q}}\right\}\\
            t_3 &= \sum_{\bmq,\bmq^{\prime}}\left(\sigma_{\bmq}\  \Tilde{\sigma}_{-\bmq^{\prime}}\right)\ \sum_{\bmq_1,\bmq_2,\bmq_1^{\prime},\bmq_2^{\prime}}\left( \frac{1}{2} \Gamma^3_{\bmq_1,\bmq_2,\bmq}\right)\left( \frac{1}{2} \Gamma^{3 *}_{\bmq_1^{\prime},\bmq_2^{\prime},\bmq^{\prime}}\right)\langle \phi_{\bmq_1}\phi_{\bmq_2}\phi_{-\bmq_1^{\prime}}\phi_{-\bmq_2^{\prime}}\rangle
        \end{align}
\end{subequations}
Now we can use the zero tadpole condition to trade 
\begin{equation*}
    \frac{1}{2}\sum_{\bmq} \frac{\Gamma^3_{\bmq,-{\bmq},\mathbf{0}}}{\Gamma^{2,\phi}_{R,q}} \leftrightarrow -\Gamma^1.
\end{equation*}
Doing this shows that $t_2=-2 t_1$. Furthermore, we can use the fact that we are taking expectation values in the Gaussian $\phi$ state to use Wick's theorem in expanding out $\langle \phi_{\bmq_1}\phi_{\bmq_2}\phi_{-\bmq_1^{\prime}}\phi_{-\bmq_2^{\prime}}\rangle$. There is one term that is $\langle \phi_{\bmq_1}\phi_{\bmq_2}\rangle\langle\phi_{-\bmq_1^{\prime}}\phi_{-\bmq_2^{\prime}}\rangle$, which gives a contribution $+t_1$ as well as two other contractions. Together, we see that the $\mathcal{O}(g^2)$ contributions proportional to $\sigma_{\bmq=\mathbf{0}}\ \Tilde{\sigma}_{\bmq=\mathbf{0}}$ all cancel! 

What is then left are the two other contractions in $t_3$. Keeping track of all the Kronecker deltas that appear and making use of the fact that all momenta are being summed over, we can write final contribution to the reduced density matrix as:
\begin{equation}
    \frac{1}{2}\sum_{\bmq}\left( \left(\sigma_{\bmq} \Tilde{\sigma}_{-\bmq}+\sigma_{-\bmq} \Tilde{\sigma}_{\bmq}\right) \left(\sum_{\bmq_1,\bmq_2}\  \frac{\left| \Gamma^3_{\bmq_1,\bmq_2,\bmq} \right|^2}{2\  \Gamma^{2,\phi}_{R, q_1} \Gamma^{2,\phi}_{R, q_2}} \right)\right)\Psi_G\left[\sigma\right]\Psi_G^*\left[\Tilde{\sigma}\right]
\end{equation}
Within our perturbative expansion, we can exponentiate this term to arrive at:
\begin{equation}
\label{eq:reducedcubidensitymatrix}
       \langle \sigma\left | \varrho_{\mathrm{red}} \right | \Tilde{\sigma}\rangle = N_{\mathrm{red}}(t)\exp\left(-\frac{1}{2}\sum_{\bmk}\left(\begin{array}{cc}\sigma_{\bmk}, & \Tilde{\sigma}_{\bmk} \end{array}\right)\mathcal{R}_k  \left(\begin{array}{c}\sigma_{-\bmk}\\ \Tilde{\sigma}_{-\bmk}\end{array}\right) \right),
\end{equation}
with
\begin{equation}
    \mathcal{R}_k = \left(\begin{array}{cc}\Gamma^{2, \sigma}_k & -\mathcal{M}_k \\ -\mathcal{M}_k &\ \Gamma^{2,\sigma *}\end{array}\right), \quad \mathcal{M}_k\equiv \sum_{\bmq_1,\bmq_2}\  \frac{\left| \Gamma^3_{\bmq_1,\bmq_2,\bmk} \right|^2}{2\  \Gamma^{2,\phi}_{R, q_1} \Gamma^{2,\phi}_{R, q_2}}
\end{equation}
The expression for the mixing term $\mathcal{M}_k$ bears a striking resemblance to what you might get when doing a unitarity cut on a diagram with two external $\sigma$ legs and a $\phi$ bubble; the mod squared of the vertex and two $\phi$ propagators. 

Hermiticity of the reduced density matrix is equivalent to the requirement that $\langle \sigma\left | \varrho_{\mathrm{red}} \right | \Tilde{\sigma}\rangle=\langle\Tilde{\sigma}\left | \varrho_{\mathrm{red}} \right | \sigma\rangle^*$, which in turn leads to: $\mathcal{R}_k = \sigma_1 \mathcal{R}_k^* \sigma_1$, where $\sigma_1$ is the Pauli matrix acting in on the  vector $(\sigma_{\bmk}, \Tilde{\sigma}_{\bmk})^T$. It's easy to check that this is true so the density matrix is in fact hermitian.

Eq.~\eqref{eq:reducedcubidensitymatrix} also shows that the reduced density matrix is Gaussian to this approximation and that it factorizes mode by mode. The former result is due to the fact that the cubic interaction is in fact linear in the system field while the factorization is a result of translational invariance.

We can use the factorization property to just focus on a single field mode, and then collect all the results together appropriately. For example, we can calculate the normalization constant for the reduced density matrix of wavenumber $\bmk$ by doing the relevant trace:
\begin{equation}
\label{eq:modenorm}
    \begin{split}
        1& = N^2_{\mathrm{red},\bmk}(t) \int \mathcal{D}^2\sigma_{\bmk}\ \langle \sigma_{\bmk}\left | \varrho_{\mathrm{red},\bmk} \right | \sigma_{\bmk}\rangle\\
        & = N^2_{\mathrm{red},\bmk}(t) \int \mathcal{D}^2\sigma_{\bmk}\ \exp\left(-2 \left(\sigma_{\bmk}\sigma_{-\bmk}\right)\left(\Gamma^{2,\sigma}_{R,k}-\mathcal{M}_k\right)\right)\\
        &=N^2_{\mathrm{red},\bmk}(t)\left(\frac{\pi}{2\left(\Gamma^{2,\sigma}_{R,k}-\mathcal{M}_k\right)}\right),
    \end{split}
\end{equation}
where the factor of $2$ in the exponent comes from combining the $\pm\bmk$ contributions, as does the squaring of the normalization.
Finally, we can compute the mode purity, $\gamma_{\bmk}$:
\begin{equation}
\label{eq:cubicmodepurity}
    \begin{split}
        \gamma_{\bmk} &=  \int \mathcal{D}^2\sigma_{\bmk}\ \mathcal{D}^2 \Tilde{\sigma}_{\bmk}\ \left |\langle \sigma_{\bmk}\left | \varrho_{\mathrm{red},\bmk} \right | \Tilde{\sigma}_{\bmk}\rangle \right |^2\\
        & = \left |N^2_{\mathrm{red},\bmk}(t)\right |^2\ \int \mathcal{D}^2\sigma_{\bmk}\ \mathcal{D}^2 \Tilde{\sigma}_{\bmk}\  \exp\left(-2\left(\begin{array}{cc}\sigma_{\bmk}, & \Tilde{\sigma}_{\bmk} \end{array}\right)\left(\mathcal{R}_k+\mathcal{R}^{\dagger}_k\right)  \left(\begin{array}{c}\sigma_{-\bmk}\\ \Tilde{\sigma}_{-\bmk}\end{array}\right)\right)\\
        &= \frac{\pi^2 \left |N^2_{\mathrm{red},\bmk}(t)\right |^2}{4\left(\left(\Gamma^{2,\sigma}_{R,k}\right)^2-\mathcal{M}_k^2\right)}=\frac{\Gamma^{2,\sigma}_{R,k}-\mathcal{M}_k}{\Gamma^{2,\sigma}_{R,k}+\mathcal{M}_k}=1-\frac{2 \mathcal{M}_k}{\Gamma^{2,\sigma}_{R,k}-\mathcal{M}_k}\simeq 1-\frac{2 \mathcal{M}_k}{\Gamma^{2,\sigma}_{R,k} },
    \end{split}
\end{equation}
where the last approximate equality is a perturbative statement. Note that $\mathcal{M}_k>0$, so the purity decreases from its initial value in the pure state, as it should.

We can see that we get the same result for the purity as in the paper, by making use of the fact that for either field, 
\begin{equation}
    \Gamma^2_{R,k} = \frac{1}{2 \left| f_k\right |^2},
\end{equation}
where $f_k$ is the relevant mode. We can solve eq.~\eqref{appeq:kernelmodes} for $\Gamma^3_{\bmq_1,\bmq_2, \bmq_3}$ 
\begin{equation}\label{eq:Gamma3Modes}
    \Gamma^3_{\bmq_1,\bmq_2, \bmq_3}(t)=\left(\frac{-2 i g}{\sqrt{V}}\right)\ \frac{1}{u_{q_1}(t)\ u_{q_2}(t)\ v_{q_3}(t)}\int_{t_0}^t dt^{\prime} \left(u_{q_1}(t^{\prime})\ u_{q_2}(t^{\prime})\ v_{q_3}(t^{\prime}) \right),
\end{equation}
with $\bmq_1+\bmq_2+\bmq_3=\mathbf{0}$ and we focus on the $\kappa_s=\kappa_t=0$ case for simplicity. Putting all of this together and making use of the fact that in the continuum limit, $\delta_{\bmq_1+\bmq_2+\bmq_3,\mathbf{0}}\rightarrow \left(2\pi\right)^3\slash V\ \delta^{(3)}\left(\bmq_1+\bmq_2+\bmq_3\right)$, we arrive at the result in our previous work.

\section{An Influence-Functional perspective}
\label{App:OpenEFT}

We could also imagine doing our calculations using the Schwinger-Keldysh path integral. The single-field description is then provided by an influence functional capturing both the Hamiltonian and diffusive-dissipative evolution of the system \cite{Breuer:2007juk}. In this Section, we present the field theory derived from this reduced description, which we refer to as the Open EFT.

\subsection{Schwinger-Keldysh formalism}

Let us express the Schwinger-Keldysh generating functional associated to our problem. This object allows us to extract to correlators of the theory. Working within the Schwinger-Keldysh contour, we double the fields $\sigma_+$ and $\sigma_-$, $\phi_+$ and $\phi_-$. It is convenient to work in the Keldysh basis where
\begin{align}\label{eq:basis}
    \sigma_r = \frac{\sigma_+ + \sigma_-}{2}, \qquad \sigma_a = \sigma_+ - \sigma_-, \qquad \mathrm{and} \qquad   \phi_r = \frac{\phi_+ + \phi_-}{2}, \qquad \phi_a = \phi_+ - \phi_-.
\end{align}
Following \cite{Salcedo:2024smn}, we construct the generating functional of the diagonal density matrix element
\begin{align}
\mathcal{Z}\left[J^\sigma_r, J^\sigma_a; J^\phi_r, J^\phi_a \right] = \int_{\Omega}^\sigma \mathcal{D}\sigma_r &\int_{\Omega}^0 \mathcal{D}\sigma_a \int_{\Omega}^\phi \mathcal{D}\phi_r \int_{\Omega}^0 \mathcal{D}\phi_a  e^{i \int \dd^4 x \sqrt{-g}\left(J^\sigma_r\sigma_a + J^\sigma_a\sigma_r + J^\phi_r\phi_a + J^\phi_a\phi_r \right)} \Bigg.\nonumber \\
&\quad e^{iS^\sigma_{0}\left[ \sigma_r, \sigma_a\right]} e^{iS^\phi_{0}\left[ \phi_r, \phi_a\right]} e^{iS_{\mathrm{int}}\left[ \sigma_r, \sigma_a, \phi_r, \phi_a\right]}.\Bigg.
\end{align}
Let us describe the various pieces of the generating functional.

\paragraph{Free theory}

The associated free parts are
\begin{equation}\label{eq:freepi}
S_{0}^{\sigma}\left[ \sigma_r, \sigma_a\right] = -\frac{1}{2}\int d^{4}x \left(\sigma_r,\sigma_a\right)\begin{pmatrix}
0 & \widehat{D}^\sigma_{A} \\ \widehat{D}^\sigma_{R} & -2i\widehat{D}^\sigma_{K}
\end{pmatrix}\begin{pmatrix}
\sigma_r \\ \sigma_a
\end{pmatrix},
\end{equation}
and 
\begin{equation}
S_{0}^{\phi}\left[ \phi_r, \phi_a\right] = -\frac{1}{2}\int d^{4}x \left(\phi_r,\phi_a\right)\begin{pmatrix}
0 & \widehat{D}^\phi_{A} \\ \widehat{D}^\phi_{R} & -2i\widehat{D}^\phi_{K}
\end{pmatrix}\begin{pmatrix}
\phi_r \\ \phi_a
\end{pmatrix},
\end{equation}
whose inverses of the matrices appearing in the above are given by the free propagators of the theory 
\begin{equation}
\begin{pmatrix}
0 & \widehat{D}^{\sigma}_{A} \\ \widehat{D}^{\sigma}_{R} & -2i\widehat{D}^{\sigma}_{K}
\end{pmatrix}\begin{pmatrix}
G^{K}_\sigma(x,y) & G^{R}_\sigma(x,y) \\  G^{A}_\sigma(x,y) & 0
\end{pmatrix}=\begin{pmatrix}
\delta(x-y) & 0 \\  0 & \delta(x-y)
\end{pmatrix}\,,
\end{equation}
and similarly on the left and for $\phi$. The full theory being unitary, the free propagators are simply given in terms of the mode functions, see e.g. \cite{Burgess:2009bs}. Explicitly, in spatial Fourier space,
\begin{align}
     G^{R}_\alpha(k; t, t') &= 2 \Imag{v_\alpha(k, t) v^*_\alpha(k, t')}  \theta(t - t') \label{eq:sigmaGR} \\
     G^{K}_\alpha(k; t, t') &  = i\Real{v_\alpha(k, t) v^*_\alpha(k, t')}, \label{eq:sigmaGK} 
\end{align}
where $\alpha = \sigma, \phi$ with $v_\alpha(k, t)$ the vacuum-normalised mode functions and the advanced component can simply be deduced from $G^A_\alpha(k; t, t') =  G^{R}_\alpha(k; t', t')$.

\paragraph{Interactions}

In the Schwinger-Keldysh basis, the theory being unitary, the interactions simply come from the contribution of each contour, that is
\begin{align}
    S_{\mathrm{int}}\left[ \sigma_r, \sigma_a, \phi_r, \phi_a\right] = S_{\mathrm{int}}\left[ \sigma_+, \sigma_+ , \phi_+, \phi_+\right] - S_{\mathrm{int}}\left[ \sigma_-, \sigma_- , \phi_-, \phi_-\right].
\end{align}
One can then use \Eq{eq:basis} to observe that unitary operators only contain odd powers of the advanced operators.

\subsection{Open EFT}

We now aim at integrating the heavy field $\phi$ and considering the single field Open EFT obtained for $\sigma$. Formally, we perform the path integral over the $\phi$ field, leading to 
\begin{align}
\mathcal{Z}\left[J^\sigma_r, J^\sigma_a \right] = \int_{\Omega}^\sigma \mathcal{D}\sigma_r \int_{\Omega}^0 \mathcal{D}\sigma_a & e^{i \int \dd^4 x \sqrt{-g}\left(J^\sigma_r\sigma_a + J^\sigma_a\sigma_r \right)}  e^{iS^\sigma_{0}\left[ \sigma_r, \sigma_a\right]} e^{iS_{\mathrm{IF}}\left[ \sigma_r, \sigma_a\right]}.\Bigg.
\end{align}
where 
\begin{align}
e^{iS_{\mathrm{IF}} \left[ \sigma_r, \sigma_a\right]} = \int_{\Omega}^\phi \mathcal{D}\phi_r \int_{\Omega}^0 \mathcal{D}\phi_a  e^{iS^\phi_{0}\left[ \phi_r, \phi_a\right]} e^{iS_{\mathrm{int}}\left[ \sigma_r, \sigma_a, \phi_r, \phi_a\right]}.\Bigg.
\end{align}
in practice, we follow a procedure similar to the one described in \cite{Proukakis:2024pua} by expanding in powers of $S_{\mathrm{int}}$, that is at second order in $g$ 
\begin{align}
    e^{iS_{\mathrm{IF}}}  = 1 + i \langle S_{\mathrm{int}} \rangle_\phi - \frac{1}{2} \langle S^2_{\mathrm{int}}
    \rangle_\phi + \cdots .
\end{align}
The theory being linear in $\phi$ and having removed the tadpole contribution, we assume for the moment that $\langle S_{\mathrm{int}} \rangle_\phi = 0$. We then identify the leading order $S_{\mathrm{IF}} \simeq \langle S^2_{\mathrm{int}} \rangle_\phi$.

\subsubsection{Mixing models}

The second order effective action is simply obtained by replacing internal $\phi$ legs by its propagators. For instance, for the linear mixing $\mu^2\sigma\phi$ generates the effective quadratic contributions
\begin{align}
    S_{\mathrm{IF}} \left[ \sigma_r, \sigma_a\right] \supset \mu^4 \int \; \dd^4 x \int \dd^4 y  &\bigg[ \sigma_r(x) \sigma_a(y)\langle \phi_a(x) \phi_r(y) \rangle_\phi +  \sigma_a(x) \sigma_r(y)\langle \phi_r(x) \phi_a(y) \rangle_\phi \nonumber \\
    +& \sigma_a(x) \sigma_a(y)\langle \phi_r(x) \phi_r(y) \rangle_\phi \bigg].
\end{align}
As expected, the influence functional is non-local and further assumptions must be performed to obtain a local open EFT. In spatial Fourier, the expression reads, 
\begin{align}\label{eq:SIFquad}
     S_{\mathrm{IF}} \left[ \sigma_r, \sigma_a\right]  &\supset \mu^4 \int \; \frac{\dd^3 \bmk}{(2\pi)^3} \int \dd t \int \dd t' \bigg[G^{A}_\phi(k; t, t')  \sigma_r(\bmk, t) \sigma_a(-\bmk,t')   \nonumber \\
    +& G^{R}_\phi(k; t, t')  \sigma_a(\bmk, t) \sigma_r(-\bmk,t') + 2 G^{K}_\phi(k; t, t')  \sigma_a(\bmk, t) \sigma_a(-\bmk,t')  \bigg],
\end{align}
where the $\phi$ field propagators are given in \Eqs{eq:sigmaGR} and \eqref{eq:sigmaGK}.

The linearity of the theory turns out to be useful to recast \Eq{eq:SIFquad} into a local EFT. We first consider the expansion  
\begin{align}\label{eq:freepropag}
    \sigma_{r}(\bmk, t') &= \sigma_{r}(\bmk, t) + (t-t') \dot{\sigma}_{r}(\bmk, t) + \frac{(t-t')^2}{2} \ddot{\sigma}_{r}(\bmk, t) + \frac{(t-t')^3}{3!} \dddot{\sigma}_{r}(\bmk, t) + \cdots \\
    &= \cos[\omega_\sigma(t-t')]\sigma_{r}(\bmk, t) + \frac{\sin[\omega_\sigma(t-t')]}{\omega_\sigma}\dot{\sigma}_{r}(\bmk, t),
\end{align}
where we used the free equations of motion $\ddot{\sigma}_{r}(\bmk, t) = - \omega_\sigma^2 \sigma_{r}(\bmk, t)$. The same hold true for $\sigma_{a}(\bmk, t')$. Performing the integral over $t'$, we obtain the local open EFT 
\begin{align}\label{eq:SIFquadfin}
     S_{\mathrm{eff}} = \int \frac{\dd^3 \bmk}{(2\pi)^3}  \int \dd t &\left[ \dot{\sigma}_r \dot{\sigma}_a  + \Delta_{12} \dot{\sigma}_r \sigma_a + (k^2 + m^2 + \Delta_{11})\sigma_r \sigma_a +  i D_{11} \sigma_a^2 + i D_{12} \sigma_a \dot{\sigma}_a \right].
\end{align}
The EFT coefficients are the obtained from 
\begin{align}
    {\Delta}_{11}&= - \mu^4   \int_{-\infty(1+i\epsilon)}^{t}\dd t'  \        \cos[\omega_\sigma(t-t')] \frac{\sin[\omega_\phi(t-t')]}{\omega_\phi} = \frac{2 \mu^4}{\omega_\sigma^2 - \omega_\phi^2} ,\label{eq:FphiphiM}\\
    {\Delta}_{12}&= \frac{\mu^4}{2}  \int_{-\infty(1+i\epsilon)}^{t}\dd t'  \        \frac{\sin[\omega_\sigma(t-t')]}{\omega_\sigma} \frac{\sin[\omega_\phi(t-t')]}{\omega_\phi} = 0,\label{eq:FphipiM}
\end{align}
and
 \begin{align}
    {D}_{11}&= 2\mu^4   \int_{-\infty(1+i\epsilon)}^{t}\dd t'  \        \cos[\omega_\sigma(t-t')] \frac{\cos[\omega_\phi(t-t')]}{2\omega_\phi} = 0,\label{eq:DphiphiM}\\
    {D}_{12}&= - \mu^4   \int_{-\infty(1+i\epsilon)}^{t}\dd t'  \        \frac{\sin[\omega_\sigma(t-t')]}{\omega_\sigma} \frac{\cos[\omega_\phi(t-t')]}{2\omega_\phi} = - \frac{1}{\omega_\phi} \frac{2 \mu^4}{\omega_\sigma^2 - \omega_\phi^2} , \label{eq:DphipiM} 
\end{align}
where we used the flat space mode functions to specify the free propagators of the $\phi$ field
\begin{align}
     G^{R}_\phi(k; t, t') &= \frac{\sin \omega_\phi (t-t')}{\omega_\phi}\theta(t-t'), \qquad \mathrm{and}\qquad
     G^{K}_\phi(k; t, t') =i\frac{\cos \omega_\phi (t-t')}{2 \omega_\phi}. 
\end{align}
\Eq{eq:SIFquadfin} is the influence functional analogue of the master equation model discussed in \cite{Colas:2022hlq} (upon taking the flat space limit of the expressions discussed there). The term controlled by $\Delta_{11}$ provides an effective mass for the theory. This is a Hamiltonian effect, which can be manifestly shown by coming back to the $+/-$ basis in which the contributions factorises between the $+$ and the $-$ branch of the path integral. On the contrary, the term $\Delta_{12}$ captures dissipation of the system $\sigma$ into the environment $\phi$. At last, the two terms controlled by $D_{11}$ and $D_{12}$ only depend on the advanced components and are pure imaginary, as it should from the non-equilibrium constraints \cite{Liu:2018kfw}. They offer a stochastic noise for the $\sigma$ sector coming from the fluctuations of the surrounding $\phi$ environment. This effect is manifestly non-Hamiltonian and cannot be expressed in terms of an in-out contour. 

If one compares the power spectrum obtained from \Eq{eq:SIFquadfin} to the ones of the two-field theory, one would realize some contributions are missing.\footnote{In the $+/-$ basis (see \cite{Chen:2017ryl} for details), the missing diagrams can be traced back to the so-called \textit{cut-in-the-middle} diagrams \cite{Senatore:2009cf} where the first vertex is taken in the $+$ and the second vertex in the $-$ contour or \textit{vice versa}.} On the top of the mass renormalization and the noise, the environment can also modify the occupation number of the state, which appear through the implementation of the $i \epsilon$ prescription. Following the approach of \cite{Salcedo:2024smn}, the effective functional appearing in \Eq{eq:SIFquadfin} really stands for
\begin{align}
&\qquad \qquad S_{\mathrm{eff}} \left[\sigma_r,\sigma_a\right] = \int \dd^4 x  \begin{pmatrix}
\sigma_r(x) & \sigma_a(x)
\end{pmatrix} \begin{pmatrix}
    0 & \left[G^R\right]^{-1} \\
    \left[G^A\right]^{-1} & \left[G^K\right]^{-1}
\end{pmatrix}  \begin{pmatrix}
\sigma_r(x) \\ \sigma_a(x)
\end{pmatrix}
\end{align}
where 
\begin{align}
 \left[G^{R}\right]^{-1} &= (\partial_0 + \epsilon)^2 - \partial_i^2 + (m^2 + \Delta_{11}), \label{eq:GReq}\\
 \left[G^{A}\right]^{-1} &= (\partial_0 - \epsilon)^2 - \partial_i^2 + (m^2 + \Delta_{11}), \label{eq:GAeq} \\
 \left[G^K\right]^{-1} &= i f(k)\epsilon + i D_{12} \partial_0.\label{eq:GKeq}
\end{align}
Solving for the Gaussian path integral then taking $\epsilon \rightarrow 0$, one recovers the flat space propagators \cite{Kamenev:2004jgp}. The function $f(k)$ controls the occupation of the state 
\begin{align}\label{eq:Ndef}
    \langle \widehat{n}_{\sigma} \rangle &= \langle \widehat{a}^\dag_\sigma \widehat{a}_\sigma \rangle =  \omega_\sigma \langle \widehat{\sigma}^2 \rangle  + \frac{\langle \widehat{\Pi}^2_\sigma \rangle}{\omega_\sigma} - 1 = \frac{f(k)}{2 \omega_\sigma} - 1.
\end{align}
The pure vacuum corresponds to $f(k) = 2 \omega_\sigma$. In the partial UV completion (two-field model), we can easily compute the occupation of the field $\langle \widehat{n}_{\sigma} \rangle$ at order $\mu^4$ from its two-point functions 
\begin{align}
    \langle \widehat{\sigma}^2 \rangle &= \frac{1}{2 \omega_\sigma} +  \frac{\mu^4 (2\omega_\sigma + \omega_\phi)}{4\omega_\sigma^3\omega_\phi (\omega_\sigma+\omega_\phi)^2}, \qquad \mathrm{and} \qquad 
    \langle \widehat{\Pi}^2_\sigma \rangle  = \frac{\omega_\sigma}{2}  - \frac{\mu^4 }{4\omega_\sigma (\omega_\sigma+\omega_\phi)^2}.
\end{align}
Consequently, 
\begin{align}\label{eq:fkfin}
    f(k) = 2 \omega_\sigma + \frac{\lambda ^2}{\omega_\phi (\omega_\phi+\omega_\sigma)^2}, \qquad \Rightarrow \qquad \langle \widehat{n}_{\sigma} \rangle = \frac{\mu^4}{2 \omega_\sigma M^3} + \mathcal{O}\left(\mu^8, \frac{1}{M^6}\right),
\end{align}
which is displaced from the pure vacuum. This change of the occupation number can be related to a change in the state's purity. The proof relies on having a Gaussian state (for which $\gamma_\bmk = 1/(4 \det \bs{\mathrm{Cov}})$ where $\det \bs{\mathrm{Cov}}$ is the determinant of the covariance matrix \cite{Serafini:2003ke}) and starting from the flat-space vacuum (for which $\langle \{\widehat{\sigma},\widehat{\Pi}_{\sigma} \}\rangle^{(0)} = 0$, where the upper index indicates the order in perturbation theory).\footnote{In particular, the detail of the interaction considered is irrelevant for the demonstration and it should hold whether we consider a linear or cubic interaction and/or a position or derivative interaction.} At second order in the system-environment coupling, the purity writes \cite{Colas:2024ysu}
\begin{align}
    \gamma_\bmk &= 1 - 4 \left[\langle \widehat{\Pi}^2_\sigma \rangle^{(0)} \langle \widehat{\sigma}^2 \rangle^{(2)} + \langle \widehat{\sigma}^2 \rangle^{(0)}  \langle \widehat{\Pi}^2_\sigma \rangle^{(2)}  - 2 \langle \{\widehat{\sigma},\widehat{\Pi}_{\sigma} \}\rangle^{(0)}  \langle \{\widehat{\sigma},\widehat{\Pi}_{\sigma} \}\rangle^{(2)} \right] + \cdots \\
    &= 1 - 2 \left[\omega_\sigma  \langle \widehat{\sigma}^2 \rangle^{(2)} + \frac{\langle \widehat{\Pi}^2_\sigma \rangle^{(2)}}{\omega_\sigma} \right] + \cdots \label{eq:connection}
\end{align}
where we replaced the zeroth order power spectra by the standard vacuum expectation values. Using the definition given in \Eq{eq:Ndef} together with the fact that $\langle \widehat{n}_{\sigma} \rangle^{(0)} = 0$, we obtain\footnote{In fact, this result is more general than the simple derivation presented here. In \cite{Colas:2024ysu} in Appendix A, it was demonstrated in full generality the direct relation between the determinant of the covariance matrix and the expectation value of the occupation number. The impact on the purity directly follows.} 
\begin{align}\label{eq:purocc}
    \gamma_\bmk = 1 - 2 \langle \widehat{n}_{\sigma} \rangle.
\end{align}
From this relation, we recover the main text result of \Eq{PurityMasterEqFflatsecular}.

As a conclusion, there are three instances where the environment $\phi$ affects the system $\sigma$
\begin{enumerate}
    \item The generation of an effective mass controlled by $\Delta_{11}$;
    \item The generation of a noise term controlled by $D_{12}$;
    \item The occupation of the system state controlled by $ f(k)$. 
\end{enumerate}
While the first two contributions do not affect the system's purity, the third one generates the main text $1/M^3$ departure from purity discussed in \Eq{PurityMasterEqFflatsecular}. A similar construction can be built for the kinetic mixing $\cL_\rint = -  \xi \, \partial_\mu \phi \, \partial^\mu \sigma $ with no new conceptual difficulty. 


\subsubsection{Cubic interactions}

\begin{figure}[tbp]
\centering
\includegraphics[width=0.55\textwidth]{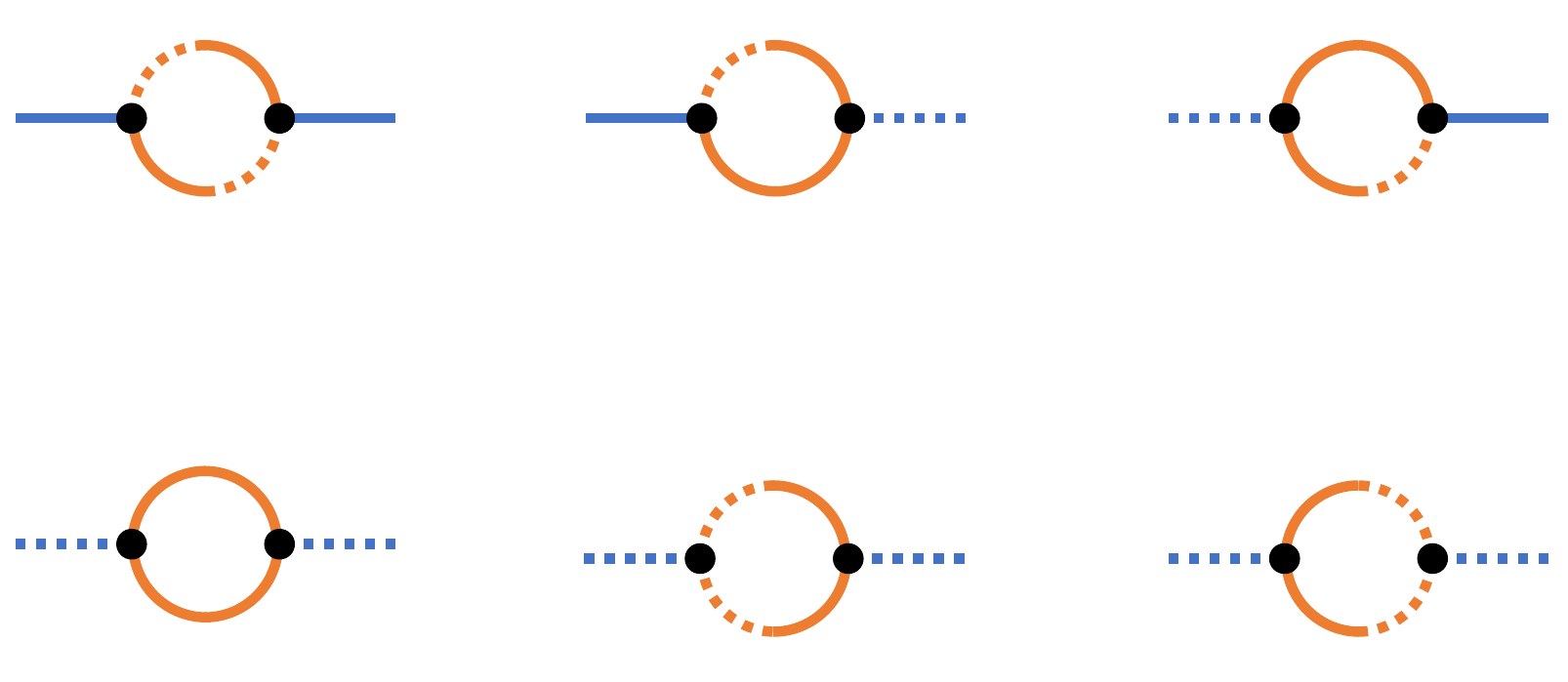}
\caption{Self-energies from the cubic interactions for the case $- \mfg \, \phi^2 \sigma$. Blue lines represent $\sigma$ while orange lines represent $\phi$; plain lines stand for retarded insertions while dashed lines stand for advanced. a) the r-r component vanishes from causality of the Green functions. b) the r-a component generates $\Sigma^A_\sigma$. c) the a-r component generates $\Sigma^R_\sigma$. d), e) and f) the a-a components generate $\Sigma^K_\sigma$.}
\label{fig:diagrams1}
\end{figure}

For an illustrative purpose, let us focus on the non-derivative cubic couplings $- \mfg \, \phi^2 \sigma$, the derivative interaction following accordingly. In this case, the self energies contributing to the second-order influence functional are given by the loop diagrams exposed in \Fig{fig:diagrams1}. We recover the known result from \cite{Kamenev:2004jgp}, that is
    \begin{align}
        \Sigma^R_\sigma(x,y) &= \mfg^2 G^R_\phi(x,y) G^K_\phi(x,y) \\
        \Sigma^A_\sigma(x,y) &=  \mfg^2 G^A_\phi(x,y) G^K_\phi(x,y) 
    \end{align}
    and
    \begin{align}\label{eq:SigmaKNL}
        \Sigma^K_\sigma(x,y) &= \mfg^2 \left\{ \left[ G^K_\phi(x,y) \right]^2 + \frac{1}{4} \left[ G^R_\phi(x,y) \right]^2  + \frac{1}{4} \left[ G^A_\phi(x,y) \right]^2 \right\}.
    \end{align}

Due to the linearity in the system's variables, the EFT structure remains the same as in the linear case given in \Eq{eq:SIFquadfin}. The only difference comes from the modification of the EFT coefficients to account for the environment non-linearities, such that 
\begin{align}
{\Delta}_{11}&= - \mfg^2   \int_{-\infty(1+i\epsilon)}^{t}\dd t'  \        \cos[\omega_\sigma(t-t')] \Sigma^R_\sigma(k;t,t') ,\\
{\Delta}_{12}&= \frac{\mfg^2}{2}  \int_{-\infty(1+i\epsilon)}^{t}\dd t'  \        \frac{\sin[\omega_\sigma(t-t')]}{\omega_\sigma} \Sigma^R_\sigma(k;t,t')  ,
\end{align}
and
\begin{align}
{D}_{11}&= 2\mfg^2   \int_{-\infty(1+i\epsilon)}^{t}\dd t'  \        \cos[\omega_\sigma(t-t')] \Sigma^K_\sigma(k;t,t') ,\\
{D}_{12}&= - \mfg^2   \int_{-\infty(1+i\epsilon)}^{t}\dd t'  \        \frac{\sin[\omega_\sigma(t-t')]}{\omega_\sigma} \Sigma^K_\sigma(k;t,t').
\end{align}
These coefficients can be explicitly obtained by performing the time integral. The difference with the previous case is that now $ \Sigma^R_\sigma(k;t,t')$ and $\Sigma^K_\sigma(k;t,t')$ are the outcome of one-loop integrals. For instance, the retarded self-energy reads
\begin{align}
     \Sigma^R_\sigma(k;t,t') = \mfg^2  \int \frac{\dd^3 \bmp}{(2\pi)^3} \int \frac{\dd^3 \bmq}{(2\pi)^3} G_\phi^R(p ,t, t') G_\phi^K(q,t, t') \delta (\bmp + \bmq + \bmk),
\end{align}
and similarly for $\Sigma^K_\sigma(k;t,t')$. We leave the explicit evaluations of these coefficients to future work.

To complete the characterization of the open EFT, we still need to specify the change in the occupation number of the state from the self-energies. Let us consider the first order Dyson equation
\begin{align}
    \bs{\mathcal{G}} = \bs{G}_0 + \bs{G}_0 \circ \bs{\Sigma} \circ \bs{G}_0 + \cdots
\end{align}
where $\circ$ represent convolution products in real space with 
\begin{align}
     \bs{G}_0 = \begin{pmatrix}
         G^K_\sigma &  G^R_\sigma \\
         G^A_\sigma & 0 
     \end{pmatrix} \qquad \mathrm{and} \qquad \bs{\Sigma} = \begin{pmatrix}
         0 &  \Sigma^A_\sigma \\
         \Sigma^R_\sigma & \Sigma^K_\sigma
     \end{pmatrix} .
\end{align}
It follows straightforwardly that the dressed propagators are given by
\begin{align}
     \mathcal{G}^{R/A}_\sigma = G^{R/A}_\sigma + G^{R/A}_\sigma \circ \Sigma^{R/A}_\sigma  \circ G^{R/A}_\sigma    
\end{align}
and
\begin{align}
    \mathcal{G}^{K}_\sigma = G^{K}_\sigma +  G^{K}_\sigma \circ \Sigma^{A}_\sigma  \circ G^{A}_\sigma +  G^{R}_\sigma \circ \Sigma^{R}_\sigma  \circ G^{K}_\sigma +   G^{R}_\sigma \circ \Sigma^{K}_\sigma \circ G^{A}_\sigma .       
\end{align}
Let us pay a closer attention to the last term. It has the exact same structure as $ G^{R}_\sigma \circ \widehat{D}^K_\sigma \circ G^{A}_\sigma$ which generates the free power spectrum. Moreover, it is easy to show in the linear case that this term captures the change in the occupation number.\footnote{For the linear case, the self-energy simply reads $\Sigma^{K}_\sigma = \mu^4 G^{K}_\phi$. In Fourier space
\begin{align}
    G^{R}_\sigma.\Sigma^{K}_\sigma. G^{A}_\sigma =  \mu^4 \left|\frac{1}{(\omega - i \epsilon)^2 - \omega_\sigma^2} \right|^2 \left[2 i \epsilon \omega_\phi \left|\frac{1}{(\omega - i \epsilon)^2 - \omega_\phi^2} \right|^2 \right]
\end{align}
where the square bracket is nothing but $G^K_\phi$ in frequency space with the vacuum prescription $f(k) = 2 \omega_\phi$. Expanding in the large mass limit $\omega_\phi \sim M \gg k, \omega$, we obtain 
\begin{align}
    \mathcal{G}^{K}_\sigma = 2 i \epsilon \left(2\omega_\sigma + \frac{\lambda^2}{M^3} \right)\left|\frac{1}{(\omega - i \epsilon)^2 - \omega_\sigma^2} \right|^2 + \cdots
\end{align}
which is indeed consistent with \Eq{eq:fkfin}} One therefore has to explicitly evaluate $G^{R}_\sigma \circ \Sigma^{K}_\sigma \circ G^{A}_\sigma$ to account for the change in the occupation number and \textit{in fine} the change in the purity of the state. 
Explicitly, the analogue of \Eq{eq:connection} leads to 
\begin{align}
    \gamma_\bmk = 1 + 2 \left\{i\omega_\sigma \left[\mathcal{G}^{K}_\sigma(k, t, t)\right]^{(2)} + \frac{i}{\omega_\sigma}  \partial_{t_1} \partial_{t_2} \left.\left[\mathcal{G}^K_\sigma(k,t_1,t_2)\right]^{(2)}\right|_{t_1,t_2\rightarrow t}\right\} + \cdots.
\end{align}
The evaluation of this expression, once expanded in the heavy mass limit, should recover \Eq{PurityMassiveEnvEqFSum2xsecular}. Again, we leave it for future work. 

\end{appendix}

\end{document}